\documentclass[sigconf, nonacm]{acmart}

\usepackage{subfigure}

\usepackage{listings}
\usepackage{xcolor}
\usepackage[ruled,linesnumbered,noend]{algorithm2e}
\usepackage[a-2b]{pdfx}
\SetKw{Def}{def}
\SetKwFor{For}{for (}{)}{}
\SetKwFor{Foreach}{foreach (}{)}{}
\SetKwFor{PForeach}{parallelly foreach(}{)}{}
\SetKw{Continue}{continue}
\SetKw{Break}{break}
\definecolor{mauve}{rgb}{0.58,0,0.82}
\definecolor{FireRed}{RGB}{227, 27, 35}     
\definecolor{OceanBlue}{RGB}{0, 125, 25}    
\definecolor{VibrantOrange}{RGB}{255, 130, 0} 
\newcommand{\revise}[1]{\textcolor{blue}{#1}}

\newenvironment{revision}
  {\color{blue}}   
  {}          
\newenvironment{revisionone}
  {\color{FireRed}}   
  {}               
  {}               
\newenvironment{revisionthree}
  {\color{VibrantOrange}}   
  {}               
\AtBeginDocument{%
  }

\usepackage{xspace}
\usepackage{makecell}
\usepackage{multirow}
\usepackage{caption}
\usepackage{subcaption}
\usepackage{amsmath}
\usepackage{hyperref}

\begin{document}

\title{$\our$: A Multi-Agent based Data System for Heterogeneous Data Analytics}

\title{$\our$: A Multi-Agent based Data Analytics System for Heterogeneous Data}



\author{Ji Sun, Guoliang Li, Peiyao Zhou, Yihui Ma, Jingzhe Xu, Yuan Li}
\affiliation{%
  \institution{Department of Computer Science, Tsinghua University}
}

\email{{sunji,liguoliang}@tsinghua.edu.cn; {22371389,22371448}@buaa.edu.cn;1120222913@bit.edu.cn;liyuan25@mails.tsinghua.edu.cn}

\renewcommand{\shortauthors}{Trovato et al.}

\newcommand{\our}{\texttt{AgenticData}\xspace}
\newcommand{\ourplus}{\texttt{AgenticData+}\xspace}
\newcommand{\spider}{\texttt{Spider-2.0-Lite}\xspace}
\newcommand{\dabstep}{\texttt{DABStep}\xspace}
\newcommand{\wiki}{\texttt{Wikipedia}\xspace}
\newcommand{\da}{\texttt{DABench}\xspace}
\newcommand{\bank}{\texttt{Real-Bank}\xspace}
\newcommand{\electric}{\texttt{Real-Grid}\xspace}

\newcommand{\smolagent}{\texttt{MultiStep}\xspace}
\newcommand{\daagent}{\texttt{Amity}\xspace}
\newcommand{\pz}{\texttt{Palimpzest}\xspace}
\newcommand{\reforce}{\texttt{ReFoRCE}\xspace}
\newcommand{\codeagent}{\texttt{CodeAgent}\xspace}
\newcommand{\agentic}{\texttt{AgenticData}\xspace}
\newcommand{\combiaml}{\texttt{CombiaML}\xspace}

\newcommand{\zeroshot}{\texttt{Zero-shot}\xspace}
\newcommand{\fewshot}{\texttt{Few-shot}\xspace}
\newcommand{\manual}{\texttt{Hand-crafted}\xspace}

\newcommand{\note}[1]{\marginpar{\small\textcolor{blue}{#1}}}

\newcommand{\hi}[1]{\vspace{.25em}\noindent\textbf{#1}\xspace}

\settopmatter{printacmref=false}

\pagestyle{plain}
\pagenumbering{roman}

\clearpage
\setcounter{page}{0}
\pagestyle{plain}
\pagenumbering{arabic}

\begin{abstract} 
Current unstructured data analytics systems heavily rely on experts to code and manage complex analysis workflows, leading to high costs and significant time consumption. To address these limitations, we propose \our, an agentic data analytics system that allows users to submit natural language (NL) queries while autonomously analyzing data from various domains, including both unstructured and structured data. Initially, \our employs a multi-agent collaboration framework to automatically transform an NL query into a semantic plan  containing relational and semantic operators. This framework  includes a data profiling agent for identifying relevant data, a semantic validation agent for iterative optimization using feedback, and a smart memory agent for managing short-term context and long-term knowledge. Additionally, we propose semantic optimization techniques to efficiently refine and execute semantic plans. We have evaluated our system, \our, on five benchmarks, and experimental results demonstrated that \our achieves superior accuracy and significantly outperforms state-of-the-art methods. \our has also been deployed in real-world data analysis applications within banks and power grid companies.
\end{abstract}


\maketitle


\vspace{-.65em}
\section{Introduction}
\vspace{-.35em}

While large language models (LLMs) have made significant strides in understanding and generating unstructured data, substantial gaps remain in their ability to perform {\it data analytics on heterogeneous data consisting of both unstructured and structured data}. Leveraging the capabilities of LLMs -- such as understanding, reasoning, planning, and generation -- holds great promise for transforming data analytics systems that manage the heterogeneous data.

Although there have been some efforts to support semantic data analytics on unstructured data~\cite{AOP,Lotus,palimpzestCIDR,TAG,DBLP:journals/corr/abs-2506-06541,DBLP:journals/pvldb/UrbanB24,DBLP:conf/cidr/UrbanB24,DocETL,unify}, these approaches have several limitations. First, some methods propose coding-based solutions that require humans to write code using semantic operators for analyzing both unstructured and structured data~\cite{Luna,UQE,palimpzestCIDR,Lotus}. However, relying on humans to manage declarative pipelines is costly, highlighting the need for natural language (NL) interfaces to facilitate data analytics across both unstructured and structured datasets. Second, some methods extend SQL to include semantic operators~\cite{UQE}, allowing for semantic analytics of structured data. Unfortunately, these methods cannot handle unstructured data due to the challenges in obtaining schemas. It is worth noting that while there are NL2SQL works~\cite{liu2025surveytexttosqlerallms,10.14778/3685800.3685905}, they are limited to SQL-based databases and lack the capability for cross-data-source analytics. Third, some methods use NL queries to perform semantic analytics on unstructured data~\cite{ZenDB}. However, they require extracting structured data from the unstructured content before analytics, making them unsuitable for scenarios where a structured schema cannot be derived.

To address these limitations, we explore semantic data analytics for both unstructured and structured data using NL queries. Recent advancements in AI agents \cite{shen2025mindmachinerisemanus, xagent2023, huang-chang-2023-towards, smolagents} have enabled these agents to write code, search for relevant information online, and handle complex tasks. However, general AI agents are not specifically designed for collaborative multi-agent systems in data analytics tasks, which results in lower-quality outcomes in these scenarios. The challenges in developing an agentic data analytics system include:

\noindent\textbf{Challenge 1:} How can we efficiently discover query-specific datasets from vast data repositories in production environments? For example, bank production systems contain thousands of diverse datasets, but many lack detailed metadata and schema documentation.

\noindent\textbf{Challenge 2:} How can we develop logically correct and valid semantic plans for an NL query? With dozens of primitive relational and semantic operators, the number of possible plans expands to trillions of variations, making the task of finding the correct solution akin to searching for a needle in a haystack.

\noindent\textbf{Challenge 3:} How can we detect plan errors and provide effective feedback to enhance plan accuracy? Since LLMs may introduce hallucinations, effectively detecting plan errors and efficiently storing and reusing these errors to provide feedback for improving plan accuracy is challenging.

To address these challenges, we propose \our, a multi-agent data analytics system on heterogeneous data. \our begins with a collaborative multi-agent planning framework that transforms an NL query into a semantic plan composed of a data profiling agent and a data planning agent. The data profiling agent is designed to thoroughly understand the data and data relationships, extract catalogs from diverse datasets, and proactively learn domain knowledge from the datasets. The planning agent is responsible for generating the semantic plan. Given the tendency of LLMs to produce errors, especially in complex plans, we design semantic and grammar validation techniques as well as smart memory management to provide essential feedback for plan correction. 


\noindent\textbf{Contribution.} In summary, we make the following contributions.

\noindent(1) We propose \our, an in-context learning-based agentic data system featuring multi-agent collaboration that can autonomously analyze multi-source heterogeneous data using NL queries (Section~\ref{sec:sys}). To our best  knowledge, this is the first agentic data analytics system designed for heterogeneous data. 

\noindent(2) We propose a data profiling agent that analyzes heterogeneous data from various perspectives and encodes the data profiles into a graph, which is then used to generate query plans 
 (Section~\ref{sec:dataprofile}).
 
\noindent(3) We introduce a multi-agent planning method that autonomously translates NL queries into high-quality plans 
 (Section~\ref{sec:plan}).

\noindent(4) We propose a novel plan validation agent to provide plan feedback (Section~\ref{sec:planvalid}) and a memory management agent to store short-term context and long-term knowledge (Section~\ref{sec:memory}).

\noindent(5) We evaluated our system on five benchmarks, and the results showed that \our achieved the highest accuracy and the lowest cost compared to state-of-the-art baselines (Section~\ref{sec:exp}). Moreover, \our has been deployed in real-world data analysis applications at banks and power grid companies.

\vspace{-.25em}\vspace{-.35em}
\section{System Overview} \label{sec:sys}
\vspace{-.35em}

\subsection{Problem Formulation}
\label{sec:def}
\vspace{-.35em}

Given a large collection of unstructured and structured data, we tackle the challenge of autonomously analyzing this heterogeneous data. Users employ natural language (NL) queries to specify their requirements for extracting insights from the underlying datasets, and the system autonomously delivers high-quality results that accurately capture the intended insights. To support data analytic queries, we design a comprehensive set of relational operators (Selection, Projection, Join, Group, Sort, Agg, Union, Intersect, Diff) as well as semantic operators (Semantic Scan, Semantic Extract, Semantic Filter, Semantic Group, Semantic Join, Semantic Classify). These semantic operators are specifically crafted for analyzing unstructured data or textual columns~\cite{unify, palimpzestCIDR}.  Additionally, we leverage LLMs to generate codes for supporting non-predefined operators. When given an NL query, we translate it into a tree-structured semantic plan composed of these operators. We then optimize and execute the semantic plan to minimize costs without compromising accuracy.

\noindent\textbf{Problem Formulation.} Given a task $\mathcal{Q}$, a set of data sets $\mathcal{D}=\{d_1,d_2,\cdots,d_n\}$, and a set of pre-defined relational and semantic operators $\mathcal{O}=\{o_1,o_2,\cdots,o_m\}$. The data profiling agent constructs a profiling graph $\mathcal{G}(\mathcal{D})$ on $\mathcal{D}$ offline. The query planning agent generates a plan $\mathcal{P}=plan(\mathcal{Q},\mathcal{G}(\mathcal{D}),\mathcal{O})$ using the data profile. Validators and memory provide the critic feedback for the plan $\mathcal{C}(\mathcal{P})$, and our method iteratively updates the plan $\mathcal{P}_{i+1}=arg\min_{\mathcal{P}}(\mathcal{C}(\mathcal{P}_i))$. The optimizer aims to reduce the execution cost and query latency of the plan, and transforms $\mathcal{P}$ into a more efficient plan. The executor then runs the optimized plan and generates the final result.

\vspace{-1em}
\subsection{System Architecture}
\label{sec:arch}
\vspace{-.35em}

To achieve autonomous data analytics capabilities, \our is powered by multiple collaborative data agents. These agents are responsible for profiling the underlying data, selecting relevant data, generating a high-quality semantic logical plan, optimizing it into an effective semantic physical plan, and executing the plan efficiently. The data profiling agent constructs data profiles for structured and unstructured data, which are used to identify relevant data for orchestrating a plan. The data planning agent generates a high-quality plan, which includes a memory agent to store both short-term information (such as error feedback) and long-term information (such as planning knowledge), along with a plan validator that provides feedback to ensure the plan's correctness. The plan optimization agent optimizes the logical plan to reduce LLM costs. The plan execution agent efficiently executes the plan.

\noindent\textbf{Data Profiling Agent (Section~\ref{sec:dataprofile}).} It analyzes both unstructured and structured data from various perspectives and generates a data profile for them. The data profile is encoded as a graph, where each node represents a data profile (including metadata, summary, embedding) for either an unstructured document or a structured table, and each edge indicates the relationships between two nodes (including data linkage and metadata relationships). 

\begin{figure}[!t]
    \centering
    \includegraphics[width=0.95\linewidth]{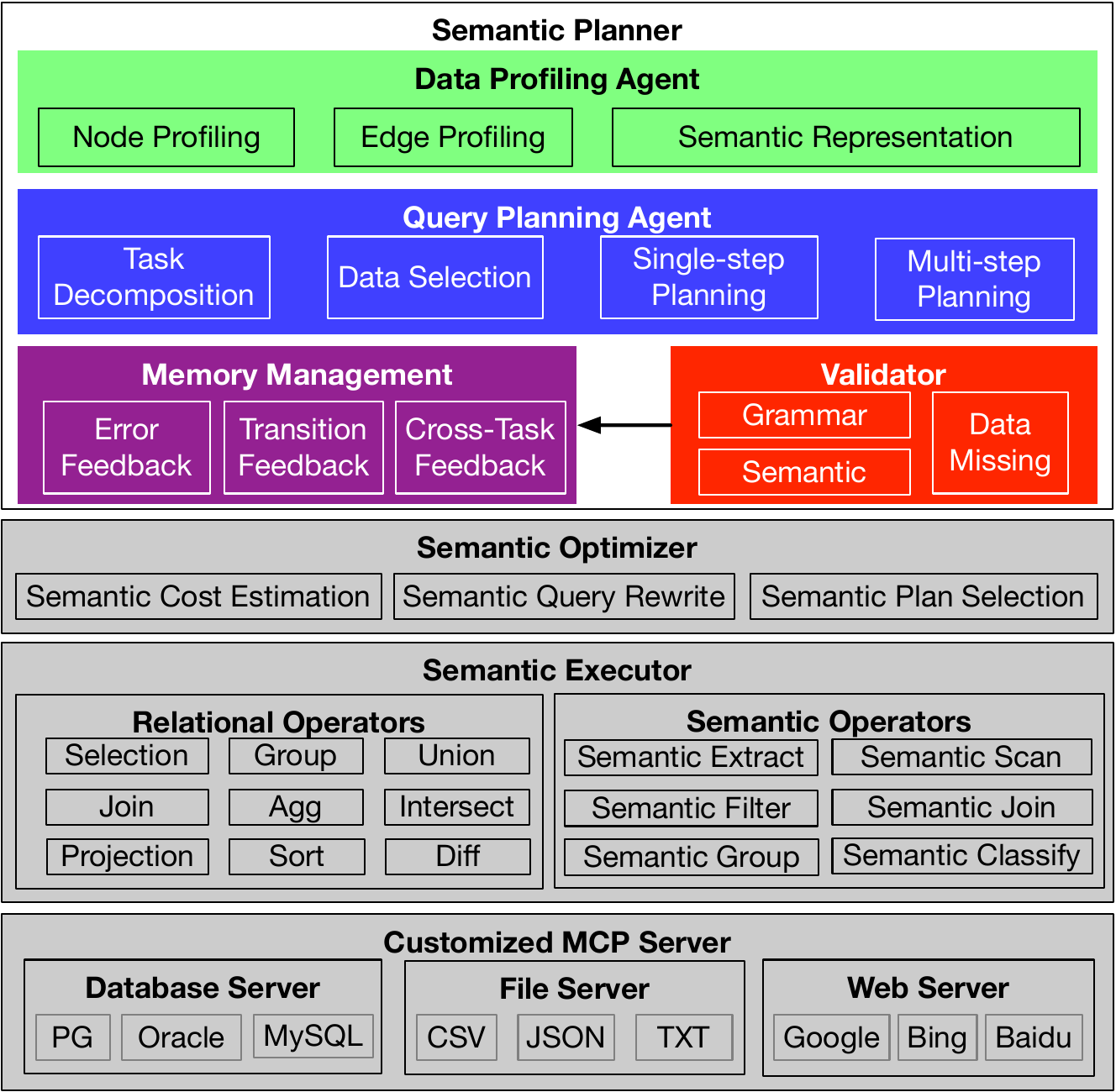}
    \vspace{-1.25em}
    \caption{Architecture of Our $\our$ System.}
    \label{fig:arch}
    \vspace{-2em}
\end{figure}

\noindent\textbf{Query Planning Agent (Section~\ref{sec:plan}).} Unlike existing NL2Data systems that generate a logical plan in a single step using sufficient prompts~\cite{10.14778/3685800.3685905, AOP}, \our employs multiple specialized agents for planning, with each agent acting as an expert in specific sub-tasks. This approach enhances performance by allowing each expert agent to focus on solving a particular aspect of the problem. The data planning agent selects relevant datasets based on the semantic requirements of the task and domain knowledge, reasoning the plan step-by-step. If there are errors, the plan validator provides feedback to enhance its correctness. 

\noindent\textbf{Plan Validator Agent (Section~\ref{sec:planvalid}).} Relying solely on LLMs to comprehend lengthy contexts can lead to flawed plans containing unexpected grammatical and logical errors, as the attention mechanism might miss key information. To identify these errors before executing the plan, we propose a plan validator agent that checks both the grammar and semantic correctness of the generated plan. If any errors are detected, the plan validator agent stores detailed error messages in memory, which are then used to inform and guide subsequent plan generation.

\noindent\textbf{Feedback-Aware Memory Management Agent (Section~\ref{sec:memory}).} To optimize memory utilization, we introduce a sophisticated mechanism designed to manage both short-term errors and long-term knowledge while coordinating its use as feedback for multiple agents. \our relies on memory for data collection and provides feedback for learning through analogy and error correction. Since all errors encountered during semantic planning are managed within memory, preventing excessive errors from overwhelming the LLMs' features poses a challenge. Unlike the conventional ReAct agent framework, which integrates all error messages into agents, \our employs a smart memory management strategy to enhance effectiveness. We record both the errors and plan traces, calculate the State Value for each plan by the Bellman Equation, and move top-k most (least) valuable plans into long-term memory. Along with the summarized common knowledge, our memory guides LLM to solve tasks more accurately and efficiently. 


\noindent\textbf{Plan Optimization Agent.} The cost of invoking LLMs to execute semantic operators is the primary expense in plan execution, making a cost-effective physical plan essential for maintaining a low-cost data analytics system. In \our, we design an optimization agent that enhances both latency and cost efficiency. We propose a three-step semantic query optimization method to reduce the cost. Initially, equivalent cost optimization rules are applied to the logical plan without degrading quality (e.g., moving the semantic operators after the relational filters, moving semantic operators that are not used by other operators to the root). Then, we optimize the join orders and reduce the plan cost using a dynamic-programming algorithm. Finally, we employ a cascading optimization approach to reduce LLM costs by initially using embeddings, followed by smaller LLMs for filtering, and finally employing larger LLMs to verify results for each semantic operator.

Specifically, we design an effective cost model for semantic plans. Note that the cost of a semantic physical plan mainly stems from LLM service charges, the typical size of input/output tokens, and the frequency of LLM invocations. In \our, we calculate the expected input size for each semantic query plan by considering the average text length derived from the data and the prompt length in the semantic operator. Likewise, we determine the expected output size based on the anticipated output of the semantic operator. The frequency of LLM invocations, influenced by the number of rows, is estimated using a novel semantic plan cardinality estimation method, which uses the data profile to identify relevant data and then applies an importance sampling method to estimate cardinality, assigning higher weights to data with greater importance. In summary, the overall cost of a semantic plan is evaluated as $\sum{Cardinality \times (|InputToken|*Fee_{in} + |OutputToken|*Fee_{out})}$.

\noindent\textbf{Plan Execution Agent.} To execute the physical plan, we use a bottom-up  execution engine for executing different operators. (1) Semantic Operators: We offer two physical implementations. The first directly invokes LLM services for each row using prompts provided by our agents. The second initially uses a vector index to search for similar content in the vector database and then invokes LLMs to verify them. (2) Relational Operators: We utilize underlying engines such as RDBMS, Spark, and Pandas DataFrame through the Model-Context Protocol (MCP), which provides a solution for integrating data access tools and managing variations across data engines. The MCP server is self-describing, enabling users to easily extend supported data sources by adding new services. 

\noindent\textbf{Remark.} {\it In this paper, we focus on the query profiling agent, query planning agent, plan validation agent, and memory management agent to generate a high-quality plan. We leverage existing techniques for plan optimization and execution, and do not claim any contributions for these two components.}

\vspace{-.5em}
\section{Data Profiling}
\label{sec:dataprofile}

\begin{figure*}[!t]\vspace{-3em}
    \centering
    \includegraphics[width=0.85\linewidth]{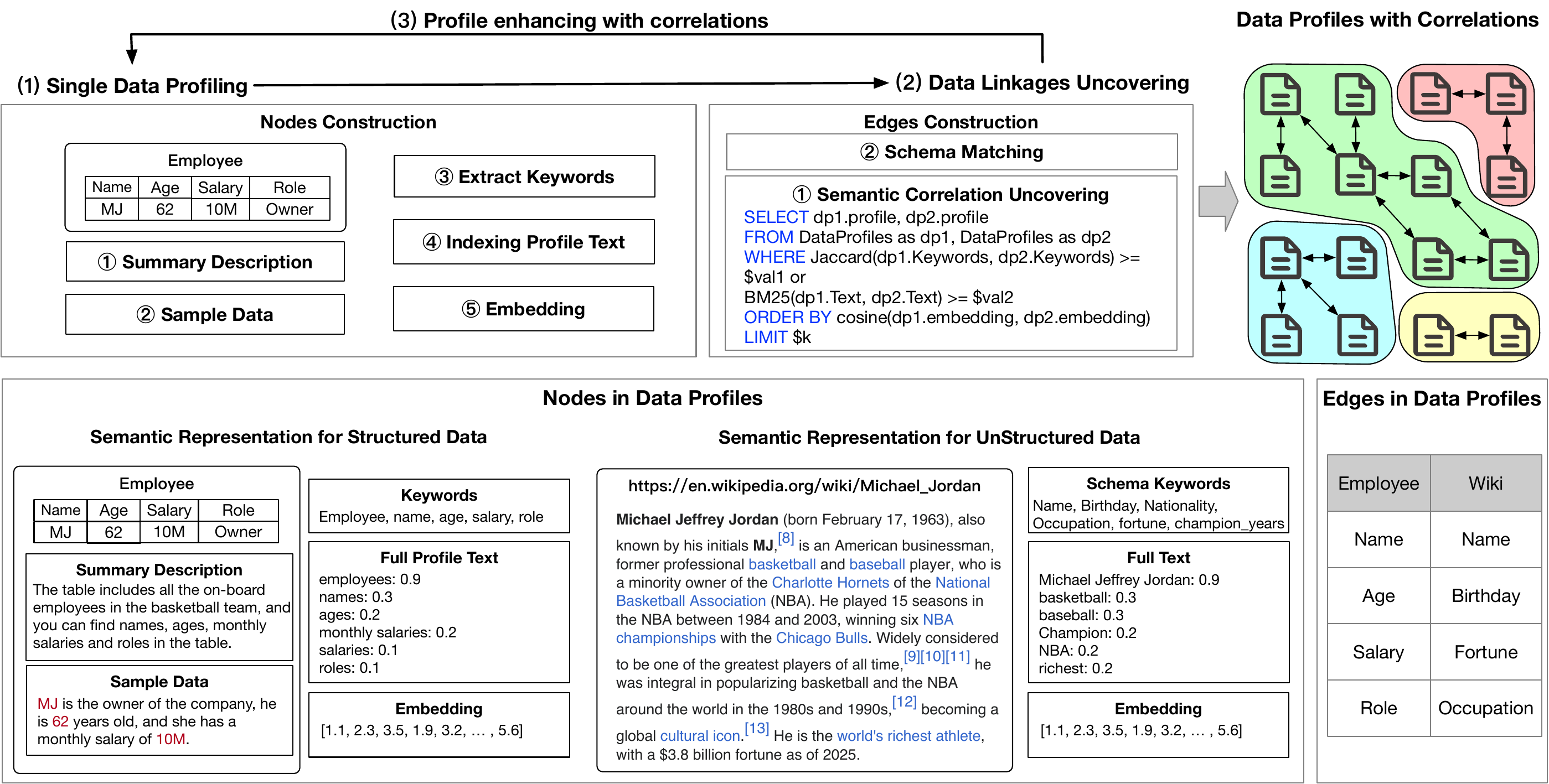}
    \vspace{-1em}
    \caption{Data Profiling.}
    \label{fig:dataprofile}
    \vspace{-1.25em}
\end{figure*}

Understanding the data schema and the relationships between datasets is fundamental to generating valid and semantically complete execution plans for a given query. Data profiling is the initial step in \our's process for solving NL queries, helping the data profiling agent select relevant data. \our analyzes the data from various perspectives and encodes the data profiles into a graph. In this graph, each node represents a data profile for either an unstructured document or a structured table, while each edge indicates the relationships between two nodes. These profiles are utilized during query planning, where the query planning agent identifies and selects data useful for orchestrating a query plan, appending the relevant profiles to the context before inputting them into the LLM for generating plans.

\vspace{-.75em}
\subsection{Data Profiling Overview}

We propose a graph-based data profiling method. Each node in the graph represents either a document for unstructured data or a table for structured data. The node profile comprises a JSON-like string that contains semantic descriptions, data schema, and data samples. The edge represents the proximity (e.g., data correlations) between two nodes. We iteratively construct the graph as shown in Figure~\ref{fig:dataprofile}. First, we generate the node profile. For unstructured data, we create semantic segments by chunking, extract summaries from unstructured datasets, and profile them with a descriptive summary. For structured data, we use table/column descriptions to extract summaries from them. Second, we convert the summary of datasets into embedding vectors, find the nearest vector neighbors and discover joinable tables or documents as correlated datasets. Third, we augment node profiling by using the neighbor nodes' profiles. To achieve this, we input the profile of a node and its neighbors' profiles into the LLM, requesting it to generate new profiles for that node and update its embeddings. We then update the edges using the updated data profiles and embedding vectors. We repeat this procedure until no further data linkages are discovered.

\vspace{-.5em}
\subsection{Node Profiling}


\subsubsection{Unstructured Data and Semi-structured Data Profiling} 

To maximize the utility of unstructured or semi-structured data, we perform schema extraction and construct a semantic catalog for their profiling. Schema extraction derives structured tables, while semantic catalog construction creates a semantic vector database from the unstructured data.

\hi{Unstructured Data With Implicit Structures.} For this type of unstructured data (e.g., tables in documents), we extract tables from the documents. Specifically, we begin by identifying structured or semi-structured data within files using tabular notations recognition or lightweight LLMs (e.g., MinerU~\cite{wang2024mineruopensourcesolutionprecise}). We then extract pertinent features from these related texts to expand the table's columns. These extracted tables are saved in separate files and profiled as normal structured files.

\hi{Unstructured Data Without Structures.} For such unstructured data (e.g., textual documents), we profile the files with a summary generated by LLMs. To support semantic search on these files, we apply a hierarchical  strategy to process them, as a single-layer technique fails to address the varied needs of RAG. For example, enhancing schema profiles requires precise and clear attribute details with minimal extraneous information, whereas planning for a particular task often demands multiple informative paragraphs that encompass extensive domain insights. To facilitate this, we initially utilize a small LLM (e.g., 7B) to segment the text into snippets. We then employ an embedding model to convert these snippets into vectors and sort them along with the original snippets in a vector database. To speed up semantic searches, we construct vector indexes for both dense and sparse vectors. After constructing the semantic snippets, we perform three steps to carry out data profiling. (1) Extract commonly used attributes in the unstructured data using NER~\cite{10.1145/3604931} techniques, and record these attributes as schema keywords; (2) Generate summary descriptions using LLMs; (3) Embed and index the summary descriptions.

\subsubsection{Structured Data Profiling}

To effectively support structured data, data profiling should extract three key insights. (1) A summarized description of the dataset helps in understanding its overall purpose and in identifying related datasets. (2) A comprehensive description of each attribute assists planners in crafting logical plans to complete tasks. (3) Identifying the datatype of each attribute facilitates the association of named entities in queries with attributes, aiding in query formation. Acquiring these detailed insights poses challenges due to missing attribute descriptions and incomplete column names. Therefore, we propose several strategies to address these issues. First, we utilize LLMs to enhance column names and descriptions, leveraging general knowledge for improved text comprehension. Second, we input sample data into LLMs to determine precise attribute meanings, which aids in identifying entities and joinable columns across datasets. Third, we use the profiles of correlated datasets to enhance the summary description and column descriptions. Lastly, we extract keywords from schema and sample data, construct keyword indexes for the profile and convert the summary into vectors via embedding.

\vspace{-.5em}
\subsection{Edge Profiling: Uncovering Data Linkages}
\vspace{-.25em}

Dataset correlations not only help enhance node profiles but also significantly improve the recall of data selection in later query planning. Data correlations are systematically created based on similarities within data profiles, which manifest in three primary forms: vector similarity, text similarity from summary descriptions, and set similarity of extracted keywords. To construct edge profiling, we propose a two-step method that first identifies all correlated datasets and then constructs edge profiles via schema matching.

In the first step, as shown in Figure~\ref{fig:dataprofile}, we perform a hybrid search to self-semantic-join all node profiles, considering three types of similarities: vector, text, and keyword. We build a set similarity index, a vector similarity index and a BM25 index to accelerate the semantic joins. In the second step, we enumerate all the data node profiles. For each node profile, we perform similarity search based on the three indexes to extract the top-$k$ closet column matching pairs. We use a default value of $k=10$. Once the data correlations are found, we can establish the data linkages for edge profiling with the matching keywords (attributes). Note that schema matching in edge profiling is used solely to identify potential join attributes, rather than directly generating matching entities (which is instead performed by running the semantic operators online).

\vspace{-.5em}
\section{Semantic Query Planning}
\label{sec:plan}
\vspace{-.35em}

In \our, semantic planning is essential for coordinating a semantic pipeline tailored to a specific NL query. Rather than relying solely on the LLMs that  can frequently make errors, \our incorporates a feedback-based framework that employs categorized grammar and semantic validators to develop the ultimate logical plan. A single planning step that explores one logical plan for a query may not yield the optimal plan, necessitating multiple steps to arrive at the best plan. Therefore, the goal of semantic planning is to develop {\it a grammatically correct and semantically valid logical plan} through multi-step reasoning. Achieving this requires {\it effective query planning} to form appropriate plan transitions, {\it comprehensive validation} for effective plan feedback, and {\it effective memory management} of historical experiences to facilitate faster plan exploration and reuse the planning knowledge of similar queries.

\vspace{-.65em}
\subsection{Overall Methodology}  \label{subsec:planoverview}
\vspace{-.35em}



\hi{Semantic Planning.} Semantic query planning can be formulated as a Markov Decision Process (MDP). Initially, the query planning agent selects relevant data profiles from vector database for the given task which are offline well prepared as discussed in Section~\ref{sec:dataprofile}. Then the query planning agent uses the selected data profiles, domain knowledge and long-term memories collected in the previous semantic planning step to generate a logical plan $\mathcal{S}_1$ that can be considered as the next step of the initial state $\mathcal{S}_0$. The long-term memory comprises common knowledge summarized from errors, good and bad plans (defined below) of similar tasks, and Q\&A context. Then the plan validation agent verifies the plan correctness, to effectively provide accurate action rewards $\mathcal{R}_{t+1}$ and comprehensive plan transition instructions. To comprehensively assess plan validation, we deploy various validators that specialize in evaluating plans from different viewpoints. If the plan validator detects errors, it puts these error messages as feedback into short-term memory. The feedback provided by the plan validator is incorporated into the state transition policy to facilitate plan evolution. Additionally, based on the memory manager's strategy, some feedback may be transferred into long-term memory for sharing across different tasks. Both short-term and long-term memory guide the LLM in query planning to transition the plan state from $\mathcal{S}_t$ to $\mathcal{S}_{t+1}$. The transition probabilities are determined by the LLM and the current contents of the memory. \our iteratively runs this process until a generated plan passes validation and is accepted.

\hi{Plan Status.} Semantic query planning involves exploring plans based on the task and memory. First, we define some key terminologies related to plan exploration traces. An \texttt{Accepted Plan} is a logical plan that has been verified by all LLM-based validators. A \texttt{Terminated Plan} is a logical plan deemed erroneous after exhausting all steps of plan exploration. A \texttt{Good Plan} is a logical plan with a State Value ranking among the top-$k$ highest, making it more likely to lead to an Accepted Plan. Conversely, a \texttt{Bad Plan} is a logical plan with a State Value ranking among the top-$k$ lowest, making it more likely to result in a Terminated Plan.

\begin{figure}[!t]
    \centering
    \hspace*{-1em}\includegraphics[width=1.08\linewidth]{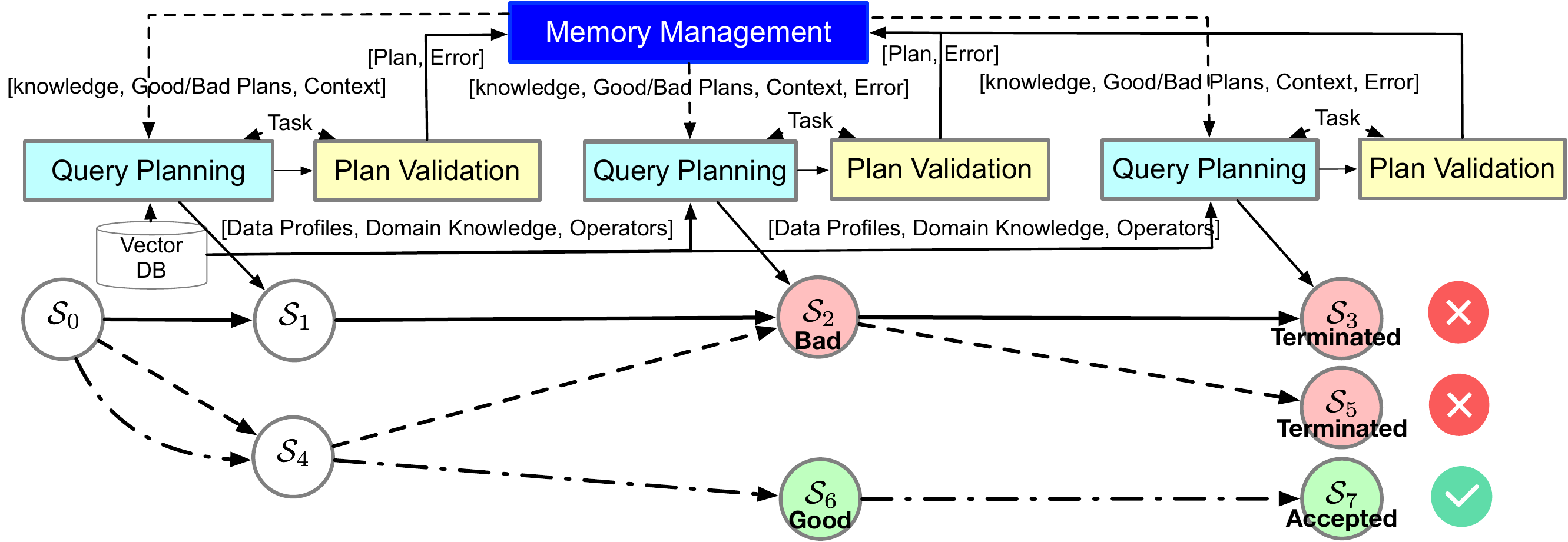}
    \vspace{-2.25em}
    \caption{Overview of Semantic Planning.}
    \label{fig:planoverview}
    \vspace{-2em}
\end{figure}

\hi{Running Example.} As Figure~\ref{fig:planoverview} shows, the whole plan exploration process can be modeled as a directed acyclic graph (DAG). In the DAG, different line styles represent different plan generation trials. \our explores the plan state of $\mathcal{S}_1$, $\mathcal{S}_2$, and $\mathcal{S}_3$ starting from the sentinel state $\mathcal{S}_0$. Since $\mathcal{S}_3$ does not pass the validation and the system determines that the current trajectory is inadequate for achieving the desired plan due to excessive step exploration or becoming cyclical, this path is terminated. Then the planning agent restarts from $\mathcal{S}_0$. At this point, the terminated plan $\mathcal{S}_3$ is taken as the bad plan by plan validator, and the memory changes the behavior of subsequent attempt of the query planning agent. In the subsequent attempt, the planner investigates $\mathcal{S}_4$, $\mathcal{S}_2$, and $\mathcal{S}_5$, but again this path is rejected because it does not pass the validation. $\mathcal{S}_5$ is put into the memory for the same reason. Moreover, the rejected states $\mathcal{S}_3$ and $\mathcal{S}_5$ both originating from $\mathcal{S}_2$, $\mathcal{S}_2$ are also labeled as bad plans. After two failed attempts, the query planning agent discovers an accept plan $\mathcal{S}_7$ in the third round, and the process terminates.

\hi{Planning Framework.} We summarize the planning framework.

\noindent Step 1: \textbf{Relevant Dataset Selection} (Section~\ref{subsec:multipleselection}). It selects relevant datasets based on data profiles constructed in Section~\ref{sec:dataprofile}.

\noindent Step 2: \textbf{Single-Step Planning} (Section~\ref{subsec:singlestep}).  It generates a plan based on the input of the query, selected datasets in Step 1, and memory information (which will be discussed in Section~\ref{sec:memory}). We also add a state for this generated plan and insert a link from the previous state  to this state in the MDP. 

\noindent Step 3: \textbf{Feedback-Aware Multi-Step Planning} (Section~\ref{subsec:multiplanning}). The planning agent assesses the quality of the plan generated in Step 2 by invoking the plan validator (see Section~\ref{sec:planvalid}), labels plan states with the highest/lowest values as good/bad plans (Section~\ref{subsec:multiplanning}), and stores the feedback in memory (see Section~\ref{sec:memory}), which will be used to guide future planning. 

(1) If the plan passes the validator (i.e., no errors), the plan is accepted, the planning is successful;

(2) Otherwise the validator reports errors. It evaluates the plan's quality and classifies it as either a good or bad plan and continues to generate a plan as follows.
\begin{enumerate}
        \item [(i)] Data-Missing Error: It re-selects the datasets by relaxing the thresholds (Step 1), and re-generates plans and states from the initial state $\mathcal{S}_0$.
        \item [(ii)] Other Errors:  It resumes plan generation from the last state (Step 2), providing feedback to the plan's quality.
\end{enumerate}

The memory dynamically manages two lists of good/bad plans, which can steer the policy of state transitions. We use LLMs with the good/bad plan feedback as a policy to conduct state transmission.

\begin{figure*}[!t]\vspace{-3em}
    \centering
    \includegraphics[width=\linewidth]{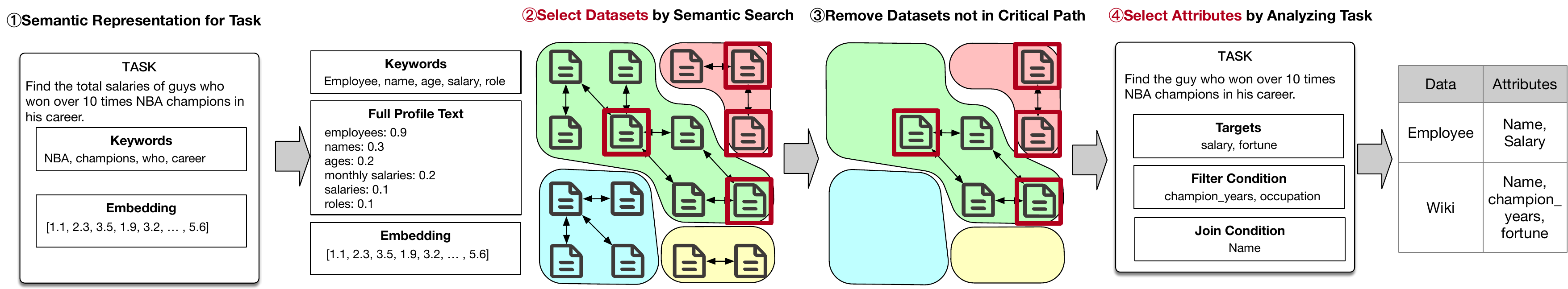}
        \vspace{-1.75em}
    \caption{Multiple Granularity Data Selection.}
    \label{fig:dataselect}
        \vspace{-1.5em}
\end{figure*}

\subsection{Multiple Granularity Data Selection} \label{subsec:multipleselection}

LLM accuracy significantly decreases when the context length surpasses a certain threshold, such as 32,000 tokens. In Figure~\ref{fig:dataselect}, we propose a multi-granularity data selection process designed to filter relevant datasets before preparing the data knowledge for the planning agent. We first identify the {\it directly relevant datasets} that have high similarity to the query, and then obtain the {\it indirectly relevant datasets} that lie along the paths between directly relevant datasets in the profile graph.

\hi{Directly Relevant Dataset Selection.} Initially, we generate the semantic embedding for the query using the same embedding method as the data profiling agent. We then perform a multi-granularity dataset search using both keywords and the embedding of the query. The search query is formalized below.

SELECT profile FROM DataProfiles 

WHERE 
       Jaccard(profile.Keywords, query.Keywords) >= \$val1 or 
       
       \hspace{3.5em}BM25(profile.Text, query.Text) >= \$val2
       
ORDER BY cosine(profile.Embedding, query.Embedding)

LIMIT \$k\\
where \$val1, \$val2, and \$k are threshold variables that control the strictness of data selection. We use default values of \$val1=0.75, \$val2=0.75, and \$k=10, relaxing these values by 10\% if data-missing errors are detected in memory. The search results of this query are directly relevant datasets. This quick process, supported by semantic indexes, examines millions of datasets in milliseconds while maintaining over 99.9\% selection recall — contrasting with the several minutes required by LLMs.

\hi{Indirectly Relevant Dataset Selection.}  Although some datasets may not be directly relevant to the query, they could still contribute by being related to the directly relevant datasets. To identify such datasets, we look for all paths between any two directly relevant datasets in the profile graph, referred to as {\it critical paths}. The datasets on these critical paths are considered {\it indirectly relevant}. We discover all critical paths using the DFS algorithm for any pair of directly relevant datasets. As illustrated in Figure~\ref{fig:dataselect}, the data selection agent identifies four directly relevant datasets and subsequently finds two indirectly relevant datasets.

\hi{Identify Projection/Selection/Join Attributes.} To align datasets with query conditions, such as projection columns, filter conditions, and join conditions, we extract attributes from data profiles. LLMs are used to extract projection and filter conditions from node profiles. Similarly, LLMs are used to determine join attributes by extracting them from edge profiles and node profile keywords/text.

Finally, the indirectly and directly relevant datasets, along with the extracted query conditions, are sent to the LLM for plan generation. This approach significantly reduces the burden of LLM inference by eliminating irrelevant data.

\vspace{-.5em}
\subsection{Single-Step Plan Generation}  \label{subsec:singlestep}
\vspace{-.25em}

In contrast to current agent-based data analytics systems that depend on the coding agent for data retrieval, \our employs a deterministic plan structured as a tree. The leaf nodes in the plan tree are data-access methods that support heterogeneous data sources and can be extended by configuring MCP servers, and the intermediate nodes are relational or semantic operators. This approach enhances plan accuracy and explainability in data analytics while being conveniently produced by LLMs without the necessity of training on vast amounts of code. The query planning agent is able to utilize the well-prepared features, including selected data profiles, task descriptions, operator details, semantic relevant domain knowledge and existing memories, to generate a comprehensive logical execution plan for addressing the task.

\hi{Query Planning.} The query planning agent takes questions, data profiles, and system memories as inputs. It first selects relevant data using multi-granularity data selection (see Section~\ref{subsec:multipleselection}) and then generates logical plans using LLMs. Generally, the query planning agent involves three steps to precisely generate a logical plan for an NL query. Initially, to better understand the domain-specific query given by users, \our seeks relevant segments from vector databases to grasp the terminology and domain knowledge. Secondly, $\our$ evaluates the data profiles and the query, employs the LLM to select relevant datasets, and obtains corresponding relevant profiles. Thirdly, the LLM takes inputs -- including the task description, selected data profiles, memories and relevant segments -- to devise a step-by-step high-level plan for tackling the query. It then reviews predefined operator definitions and formulates a comprehensive tree-structured logical plan with detailed operator specifications. We build semantic operators for unstructured data and relational operators for structured data. The system can then add more semantic operators to the text columns of structured data if feedback from validators indicates they are needed. For instance, if the data formats of intermediate data do not match, validators will detect grammar and semantic errors. Based on this feedback, we construct instructions to guide agents in adding another semantic extract operator to align the data format.



\begin{figure*}[!t]\vspace{-2em}
    \centering
    \begin{minipage}[t]{0.56\textwidth}
        \centering
        \includegraphics[width=1.05\textwidth]{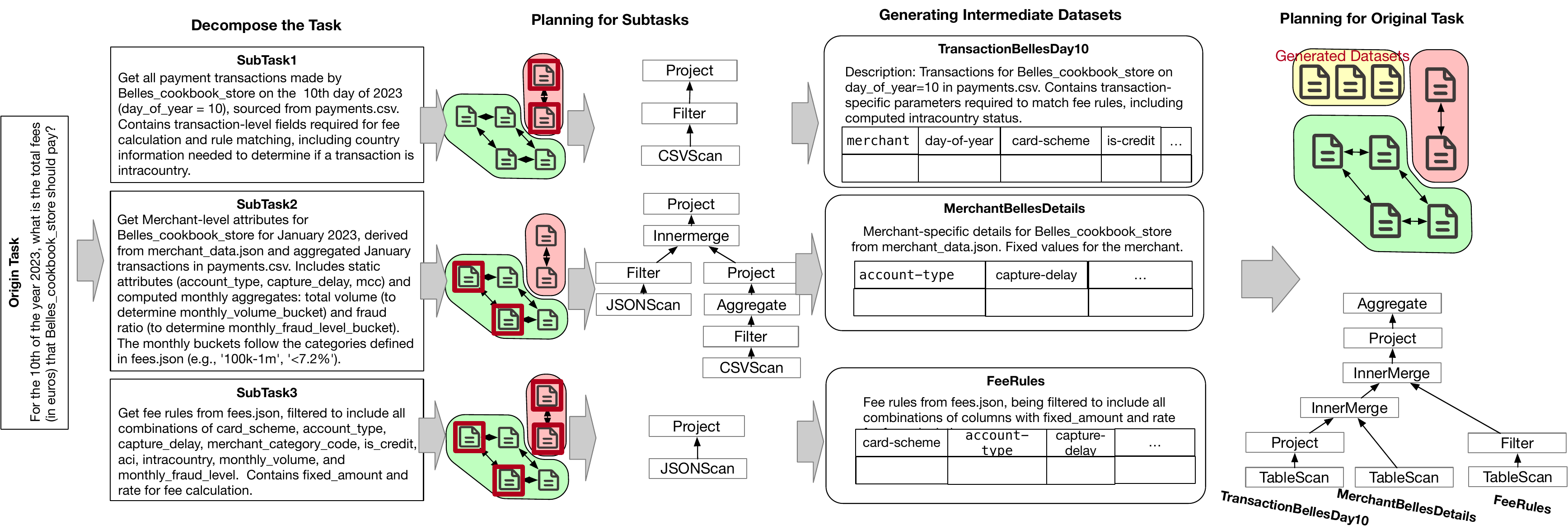}\vspace{-1em}
        \caption{Complex Query Planning: Task Decomposition and Merging.}\label{fig:bigplan}
    \end{minipage}
    \hfill 
    \begin{minipage}[t]{0.43\textwidth}
        \centering
        \includegraphics[width=\textwidth]{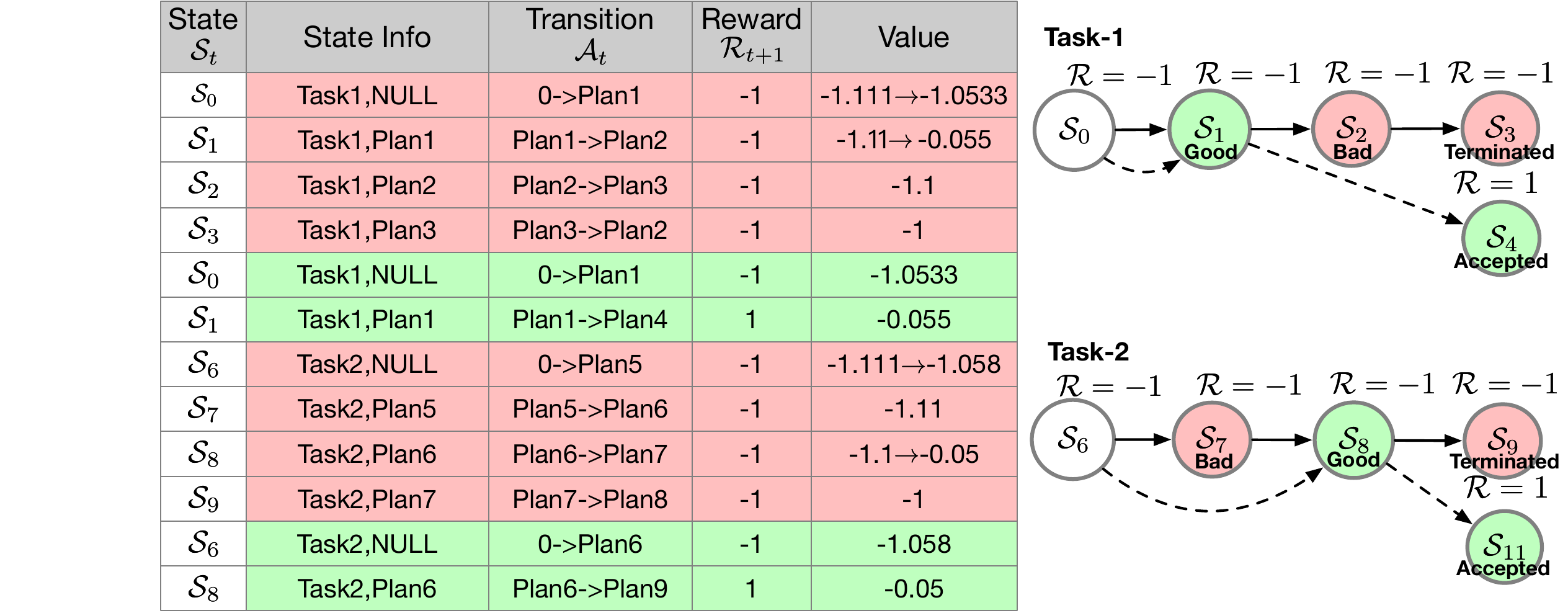}\vspace{-1em}
        \caption{Example of Reward and Value Computation.}\label{fig:values}
    \end{minipage}
    \vspace{-1.5em}
\end{figure*}

\hi{Complex Query Decomposition.} The success rate decreases as planning steps increase significantly with complex tasks, as shown in Section~\ref{sec:exp}. This decline occurs because the plan involves more data selections and an exceptionally lengthy logical plan, which can exceed 1,000 lines. In such situations, it is challenging for the LLM to fully comprehend all the information conveyed by the data/query and to formulate an accurate plan given the complexity of the query.

To address this issue, we decompose the complex query into broadly independent sub-tasks, allowing each to be logically self-contained. This approach enables \our to focus on smaller tasks and portions of datasets. There are two challenges in implementing this divide-and-conquer strategy in complex task planning. The first is ensuring that all sub-tasks are essential components of the main query and remain integral after merging. The second is ensuring that the sub-plans of sub-tasks can be easily merged.

To tackle these challenges, we first use the LLM to break down the complex query with a constraint prompt emphasizing integrity and necessity, and adopt a validator to ensure the precision of task decomposition. The prompt is constructed by the given task, data profiles, validator's suggestions and some few-shot rules. Second, we execute the planning-feedback-execution process for each sub-task and generate each sub-task's results as complementary datasets. The intermediate results are materialized for data profiling purposes, and they can be stored on disk for scalability.
As shown in Figure~\ref{fig:bigplan}, each sub-task undergoes the complete agentic workflow, ensuring accuracy through validators and feedback. Third, instead of feeding sub-plans into the LLM to generate the final logical plan for the original task, \our executes sub-plans, derives sub-task results, and extends datasets with these well-prepared results. The query planning agent then takes the generated complementary datasets and the query as input, and calls the planning agent to generate a logical plan, significantly reducing the effort required for planning complex tasks.

\vspace{-.25em}
\vspace{-.25em}
\subsection{Multi-Step Planning} \label{subsec:multiplanning}
\vspace{-.25em}

\hi{Plan Exploration.} To reduce the steps of plan exploration, we formulate the problem as a reward optimization problem. The state $\mathcal{S}$ is a logical plan explored by the query planning. The action $\mathcal{A}$ represents a transitioned subsequent plan. The policy $\pi_m(\mathcal{S})$ represents probabilities of Step 2 for generating various next logical plans, and the $\pi_m(\mathcal{S})$ can be tuned by controlling the content of memory. The reward is derived from validations in Step 3 and is associated with an action, $\mathcal{R}(\mathcal{S}, \mathcal{A})$. The reward is defined as -1 for an invalid transition plan, and 1 for a valid one. 


Our objective is to maximize the total reward of the plan exploration for each task by identifying an optimal state transition policy $\pi_m$, i.e., $\max_{\pi_m}\sum_0^{N}\mathcal{R}_{t+1}(\mathcal{S}_t, \mathcal{A}_t)$, 
where $\mathcal{S}_0 = \varnothing$, and $N$ is the total steps for exploring plans for a query. Since the policy $\pi_m$ consists of both the LLM and memory, and tuning the LLM is costly and unstable, we construct a memory to guide query planning. Generally, we use Bellman equation to assess future value for each plan state, and select actions based on the values to address the MDP optimization problem. However, due to the inability to acquire all possible transition rewards, the values must be updated incrementally. In general, to balance exploitation with exploration, we implement an in-context reinforcement learning approach to tackle the issue, and the transition policy to be optimized is the content of good/bad plans (defined in Section~\ref{subsec:planoverview}) in the memory. Next we discuss how to select good/bad plans to steer the plan exploration.

\hi{State Value Function.} The value of each transition in the plan exploration process is calculated using the Bellman Equation. Whenever an exploration process is terminated (failed or successful), we store the state and transition of each step to the memory. From the last state to the first, we compute the value using the current reward and the values of previous states. And at each step, we find the <state, transition> with similar states in the same task group (considering they are in the same state $\mathcal{S}$), and average the value of them. The calculation function is $v(\mathcal{S}_t) = \mathbb{E}[\mathcal{R}_{t+1} + \gamma v(\mathcal{S}_{t+1})]$, 
where $\gamma$ is a parameter used to balance the last-step value and the reward value in order to compute the next-step value. After we obtain the value $v(\mathcal{S})$ of state $\mathcal{S}$, we select the high-valued and low-valued states as good plans and bad plans respectively. In this way, the planning agent in \our not only receives short-term feedback but also learns which logical plans are worth pursuing and which are not. This significantly enhances both the efficiency and accuracy of plan generation.


Figure~\ref{fig:values} shows an example. For task group 1, there are two trajectories. The first  ($\mathcal{S}_0\xrightarrow{}\mathcal{S}_3$) is terminated because a cycle is found, and the second  ($\mathcal{S}_0\xrightarrow{}\mathcal{S}_4$) successfully generates a valid logical plan. We first compute the values for the first trajectory. The value of state $\mathcal{S}_3$ is -1 because $\mathcal{R}_{t+1}$ is -1 and $v(\mathcal{S}_{t+1})$ is unknown (calculated as 0). The value of state $\mathcal{S}_2$ is calculated by $\mathcal{R}_3+0.1*v(\mathcal{S}_3)$ with $\lambda=0.1$, and it is $-1+(-1)*0.1=-1.1$. Similarly, the value of $\mathcal{S}_1$ is -1.11 and the value of  $\mathcal{S}_0$ is -1.111 for the first plan exploration trajectory. Then we compute the value for the second trajectory, and the value of $\mathcal{S}_1$ becomes the average of values of $\mathcal{R}_4+0.1*v(\mathcal{S}_4)$ and  $\mathcal{R}_2+0.1*v(\mathcal{S}_2)$, and it is $((1)+(-1+0.1*-1.1)) / 2 = -0.055$. Similarly, the value of $\mathcal{S}_0$ is computed as the average of new value and the old one of $\mathcal{R}_1+0.1*v(\mathcal{S}_1)$, and it is $((-1-0.1*0.055)+(-1-0.1*1.11)) / 2 = -1.0533$. At last, we also update values of all states to the latest values. After we obtain the value $v(\mathcal{S})$ of state $\mathcal{S}$, we select the high-valued and low-valued states as good plans and bad plans respectively. Since the state values of Plan1 ($\mathcal{S}_1$) and Plan6($\mathcal{S}_8$) are -0.055 and -0.05 respectively, and they are chose as the good plans, while the state values of Plan2($\mathcal{S}_2$) and Plan5($\mathcal{S}_7$) are -1.1 and -1.11 respectively, and they are bad plans.

\vspace{-.75em}
\section{Plan Validation}
\label{sec:planvalid}
\vspace{-.25em}

Using a single-step approach with LLMs can cause hallucination issues, especially in extensive contexts. Therefore,  \our incorporates feedback into its memory to perform in-context reinforcement learning. By integrating feedback, \our  avoids repeatedly using bad plans identified in previous steps and instead favors good plans. This approach helps guide plan exploration and optimizes both the efficiency and accuracy of plan generation.  Given the high cost associated with executing plans, most errors are identified through plan validation. Thus, plan validation must be precise and comprehensive to detect these errors. Considering the various types of errors in logical plans generated by LLMs, we design error detectors to evaluate both the grammatical and semantic accuracy of these plans. These detectors can be broadly categorized into three types: semantic errors, data errors, and grammar errors.


\vspace{-.75em}
\subsection{Semantic Validation}
\label{sec:planvalid:semantic}
\vspace{-.25em}

Semantic validation aims to check the logic consistency between generated plan and the query, and  errors found in this validation stage are called semantic errors. In order to ensure the precision of the semantic error detection, we adopt specialized LLM-based error detectors to check the plan from various perspectives.

\noindent\textbf{Logic Error Detector.} This detector assesses whether the plan logically misinterprets the query. For example, if a query requires retrieving the top-$K$ results but the plan lacks a sorting operation, that constitutes a plan logic error. We propose a lightweight model using few-shot examples to detect logic errors in the generated plan by comparing it with the query, and the query is rewritten according to the relevant knowledge extracted from vector database. The LLM can then provide conclusions on error detection and offer suggestions for fixing the error. Specifically, the logic error validator uses a task description, operator descriptions, and the plan to detect logical errors. A plan represents the self-contained logic for solving a data analytic task, and an LLM can determine whether the plan's logic correctly addresses all the task's requirements. If there is a mismatch between the plan's logic and the task, the LLM identifies the missing operators based on the operator descriptions.

\noindent\textbf{Task Decomposition Error Detector.} This detector aims to check the necessity and completeness of decomposed sub-tasks. For instance, if a query requiring computing "total fees" lacks payments sub-tasks, the task decomposition is incomplete and this is a task decomposition error. We first reconstruct the original task using the sub-tasks with an LLM and compare whether the constructed task is equivalent to the original. To enhance the robustness of error detection, we first extract targets, conditions, and expressions from the task using the LLM. We then encode them into vectors (e.g., one-hot encoding for operators and attributes, embeddings for operands) and use the semantic metrics to assess similarity. If any obvious missing or redundant objects (data, operators, conditions, outputs) are identified, we mark them directly; otherwise, we rely on the LLM for final validation, and the LLM would suggest to update some sub-tasks.


\vspace{-.5em}
\subsection{Data Validation}
\label{sec:planvalid:data}
\vspace{-.25em}

Data validation aims to find the absent datasets in the logical plan, and the errors found during this validation are called data errors.

There are two types of data missing errors in the generated plan. Direct data missing errors occur when: (i) outputs (columns) are defective because the necessary information was not accessed by the plan, and (ii) filter/join conditions (rows) are defective due to missing data. Indirect data missing errors arise when direct data has been accessed, but the datasets required for joining are missing. Traditional matching techniques struggle with detecting these data missing errors because there are no explicit or formalized attribute blocks in NL queries, and many outputs are generated through complex transformation pipelines. We address this issue by taking into account the data transformation pipeline within the plan, the intents of the NL task, and the abstract contents of the data.

\noindent\textbf{Direct Data Missing Detector.} This detector evaluates whether all attributes of data that are directly required by tasks are present in the generated logical plan, relying on the outcomes of semantic validation. The direct data missing detector analyzes the error messages from semantic validators to extract dataset descriptions. If datasets are found to be missing, we amend the data profiles and choose more suitable datasets for query planning.

\noindent\textbf{Indirect Data Missing Detector.} This detector checks if additional data is required to connect existing datasets to fulfill the query requirement. For example, an incorrect join path might necessitate additional data to bridge gaps between datasets. Since missing the joined data often causes semantic errors, we first check the results of semantic validation. If the logic error is detected, we extract the error messages and analyze whether it is related to the error connections between datasets. We can also resolve related columns from the errors, and locate the missing joined data by revising the data profile graph.

\vspace{-.5em}
\subsection{Grammar Validation}
\label{sec:planvalid:grammar}
\vspace{-.25em}

Grammar validity is the basic requirement for the generated plan, and the executor will raise exceptions if the plan has grammar errors. However, executing a plan with grammar errors brings large overhead to plan exploration, and thus we rely on grammar validators to figure out grammar errors beforehand.

\noindent\textbf{Schema Error Detector.} This detector determines whether operators access non-existent attributes or if attributes are misused. For example, LLM may fabricate some non-exist columns. In order to check the column names on all operators in the plan without actually processing the data, we conduct pseudo-execution on datasets schema to derive output result schema in each plan node, and check whether the column names used in each operator exist in the data passed from the upstream operators.

\noindent\textbf{Tools Error Detector.} This detector ensures that the tools are correctly invoked, and the parameters (such as SQL queries for database scans or code for a Python interpreter) are appropriate and complete. For SQL queries and Python code, we adopt specialized tools (e.g., SQL EXPLAIN for PostgreSQL, or Pylint for Python code) to verify the grammar of SQL and Python. For other tools, we check the tool errors by comparing the tool name and parameters in the generated plan node and the MCP server.

\vspace{-.5em}\vspace{-.25em}
\section{Agentic Memory Management}
\label{sec:memory}
\vspace{-.25em}


As a multi-agent system, \our relies on memory for collecting validation errors and providing feedback to facilitate learning through analogy and error correction. All errors encountered during the semantic planning procedures are stored in memory, and it is rather challenging to effectively store and utilize this information. First, the volume of errors and context can become exceedingly large, with each query potentially producing around 10 entries (or 10,000 tokens). The memory can inflate to millions of records within an hour, so we must prevent excessive errors from overwhelming the LLM.  Second, exploring plans can involve numerous steps and may even lead to cycles without producing successful plans. Therefore, we  design a strategy to predict the value of each state transition and prune unproductive paths promptly. Third, different agents are responsible for addressing different errors, so efficient memory utilization requires proper routing of the memory.

\begin{figure}[!t]\vspace{-2em}
    \centering
    \includegraphics[width=0.8\linewidth]{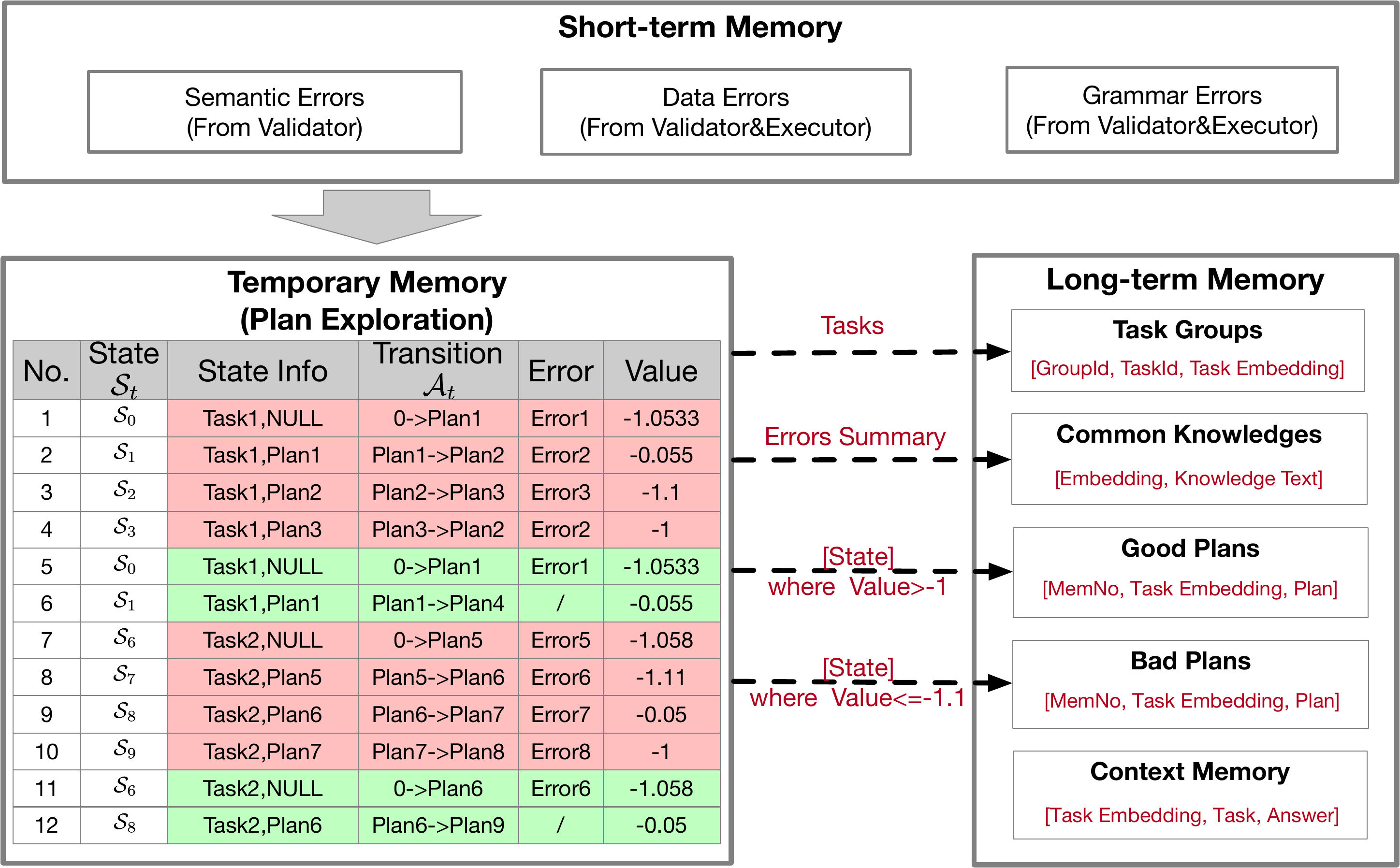}    \vspace{-1em}
    \caption{Memory Management}
    \vspace{-2em}
    \label{fig:memory}
\end{figure}

Therefore, unlike the traditional ReAct agent framework, which places all error messages into agents, \our employs a smart memory management approach to enhance feedback effectiveness. The agentic memory in \our comprises three main components: short-term memory, temporary memory, and long-term memory as shown in Figure~\ref{fig:memory}. Short-term memory stores three types of errors to provide quick feedback to the query planning agent. Temporary memory holds all planning contexts, such as successful and unsuccessful state transitions, to guide task-specific planning. Additionally, we summarize contexts from the temporary memory (e.g., good/bad plans, summarized common knowledge) and periodically copy them into long-term memory to assist with cross-task planning (e.g., upon completing the planning of a task).


\vspace{-.5em}\vspace{-.25em}
\subsection{Error Feedback in Short-Term Memory} 
\vspace{-.25em}

To effectively calibrate the plan and correct errors, short-term memory capacity is limited to holding only top-3 recent memories generated by the system for each task without sharing across plan exploration processes, and all these memories are provided to agents at each step. Short-term memory comprises three containers for different types of errors: data errors, semantic errors, and grammatical errors, which are produced by either the validation agent or the plan executor. It sends data errors to the data profiling agent to refine data profiles, directs semantic errors to the query planning agent to adjust the planning logic, and forwards grammatical errors to the query planning agent for tool parameter calibration, as well as for refining SQL queries or LLM-generated code.

\vspace{-.5em}\vspace{-.25em}
\subsection{Transition Feedback in Temporary Memory} 
\vspace{-.25em}

The temporary memory is designed to manage all planning contexts, capturing both successful and unsuccessful transitions, while also establishing long-term general knowledge. This enables \our to leverage past experiences to enhance the efficiency of plan exploration. The errors collected from the plan exploration of tasks are copied into temporary memory in the form of <state, transition, errors>, where the state is the previous plan $\mathcal{S}_t$, and the transition is the newly generated plan $\mathcal{S}_{t+1}$, and the errors are the validation feedbacks of $\mathcal{S}_{t+1}$. In temporary memory, the content is clustered by tasks, and the state transitions involved in plan exploration for a particular task are arranged sequentially. Temporary memory involves processes such as summarizing common knowledge, computing values, and exploiting and pruning transitions. Initially, \textit{common knowledge summarization} groups the errors based on classification and embedding similarity, filters high-frequency groups, and uses LLMs to summarize each group. Next, \textit{value computation} employs the Bellman Equation to calculate the value of each transition through plan exploration rewards. Lastly, \textit{transition exploitation and pruning} differentiates between good and bad plan states based on the calculated values, allowing for the reuse of successful actions and the elimination of ineffective ones in future similar tasks. 

\vspace{-.5em}\vspace{-.25em}
\subsection{Cross-Task Feedback in Long-Term Memory} 
\label{sec:longmemory}
\vspace{-.25em}

To effectively reuse the past experiences during plan exploration, we build a long-term memory. This is a real-time semantic memory built inside a vector database. We create five tables to manage various types of long-term memories, including \texttt{Task Groups}, \texttt{Common Knowledge}, \texttt{Good plans}, \texttt{Bad plans}, and \texttt{Context Memory}. We group tasks by semantics. The {\it Task Groups} section maps tasks to their group IDs, and the long-term memories are shared within each task group. The {\it Common Knowledge} section comprises summarized history knowledge texts and their embeddings. The {\it Good/Bad plan} sections include features of plan encoding, error embedding and acceptable/declined plans for each task group. The {\it Context Memory} section generally stores all existing tasks and their answers for co-reference resolution and associated task details, including task embedding and task descriptions. In these tables, we build semantic indexes (e.g., vector indexes) to accelerate the memory searching. In the query planning agent, the task will search the long-term memory and get the shared memories for more effective planning. Specifically, we first convert the task to embedding vectors, and select the nearest task group. Then we extract context memory from relevant questions and enhance the task using history results in the session. We then select common knowledge which may improve the planning via semantic similarity search between task and common knowledge. Finally, we select good/bad plans of similar tasks to steer the plan exploration when query planning. 

The interfaces include initiation, insertion, deletion and retrieval.

\hi{Initialization.} Given a set of NL queries, we first replace entity values with entity names (e.g., US becomes Country) using the LLM and data profiles for each query. We then extract the query skeleton as a query template by removing the value from the query plan. Since the query template defines the task's purpose, logic, and expected outcome, and similar templates often share common characteristics, we group tasks based on these template similarities. Specifically, we convert the query into an embedding and identify nearby group centroids with a similarity distance greater than a specified threshold (0.75 by default). If a match is found, we add the query to the closest group and recalculate the centroids; otherwise, we create a new group for the query. Based on this task classification, we build the \texttt{Task Groups}. This approach helps reduce the likelihood of grouping far-apart queries together.


\hi{Insertion. } Given a task and a memory (e.g., error summary, good plan, bad plan, etc.), we convert the whole task to an embedding vector, and calculate the closest group for the query by comparing the similarity of task embedding and the query target embedding of each group. Next, we insert the memory into that group.

\hi{Deletion. } Only the good/bad plans are removed when they are no longer significant. The temporary memory updates the state values constantly, and when a good/bad state is evicted from the top-k highest/lowest lists, its memory number is sent to the long-term memory. Then the plan will be removed from long-term memory.

\hi{Retrieval. } For a new task, we first extract the target of the query and convert the target and whole task to embedding vectors, and then we fetch the top-$k$ nearest memories by similarity of task embedding from the closest task group.

\begin{table}[!t]{\footnotesize
    \centering   \vspace{-3em}
    \caption{Benchmarks.}\vspace{-1.5em}
    \begin{tabular}{|c|c|c|c|}
    \hline
       Benchmark & \#Datasets & \#MaxRows & \#Questions \\\hline
       \dabstep  & 7 & 56,310 & 460 \\\hline
       \da & 68 & 150,000 & 311 \\\hline
       \spider & 13,963 & 1,751,216 & 547 \\\hline
       \bank  & 9 & 1,476 & 50 \\\hline
       \wiki  & 1000 & 408 & 100 \\\hline
    \end{tabular}
    \label{tab:datasets}
    }\vspace{-1.75em}
\end{table}

\vspace{-0.5em}
\section{Experiments} \label{sec:exp}
\vspace{-.5em}
\subsection{Experimental Settings}\label{sec:exp:setting}
\vspace{-.25em}

\noindent\textbf{Datasets.}
We have conducted extensive experiments on three popular benchmarks and two real datasets, including \dabstep~\cite{dabstep,DABStepURL}, \spider~\cite{lei2025spider20evaluatinglanguage,spider2}), \da~\cite{hu2024infiagentdabenchevaluatingagentsdata,DABench}, \wiki~\cite{WIKI}, and \bank~\cite{RealBank}. \dabstep contained 450 tasks from real-world challenges in the financial field, with heterogeneous data formats. \da contained advanced data analysis questions on 55 structured tables, and the tasks included data preprocessing, feature engineering, outlier detection, correlation analysis and machine learning.  \spider contained 547 challenging real-world data analytic problems derived from enterprise-level database use cases. \bank was a real-customer dataset from a bank that contained 50 frequently data analytic questions for banking business analysis, and a masked dataset of real deposits and loans. \wiki contained 1000 randomly selected web pages and 100 queries from Stack Exchange on these pages. Table~\ref{tab:datasets} showed the statistics of these benchmarks.

\noindent\textbf{Baselines.} We compared with state-of-the-art baselines listed on the benchmark {\it leaderboards}. \daagent was the top-1 method on \dabstep that designed a generic multistep agent framework that relied on coding ability of LLMs and a Python interpreter to solve data analytics tasks. \smolagent was the top-2 method (with codes) on \dabstep which designed a generic multi-agent technique that added a planner agent before calling the coding multistep agent. \codeagent~\cite{zhang2024codeagentenhancingcodegeneration} built a ReAct based agent framework to write code and provided a sandbox for safely running Python codes, which was used by \da to test varying agents. \reforce~\cite{deng2025reforcetexttosqlagentselfrefinement} was an NL2SQL agent method based on majority consensus, and it was the top-1 method on \spider. \pz~\cite{palimpzestCIDR} was a semantic operator system which relied on human-crafted semantic extraction pipelines. \zeroshot processed the queries using single-round prompting. \fewshot provided the LLM with hand-crafted task solutions in natural language and generated plans in a single round. \manual directly used hand-crafted plans, which required approximately two months of effort from a human expert. To reduce the interference of outliers, we repeated to run all the experiments for five times and report the median results.

\noindent\textbf{Environment.} LLMs and embedding models were executed on a machine equipped with 192 INTEL(R) XEON(R) 8558 CPU cores, 8 NVIDIA H200 GPUs, and 2TB of memory. All systems were run on laptops featuring an Apple M3 chip and 128GB of memory.

\vspace{-.65em}
\subsection{Overall Accuracy Comparison}
\vspace{-.35em}

\begin{figure*}[!t]\vspace{-3em}
    \centering
    \subfigure[$\dabstep$-easy]{
        \includegraphics[width=0.15\linewidth]{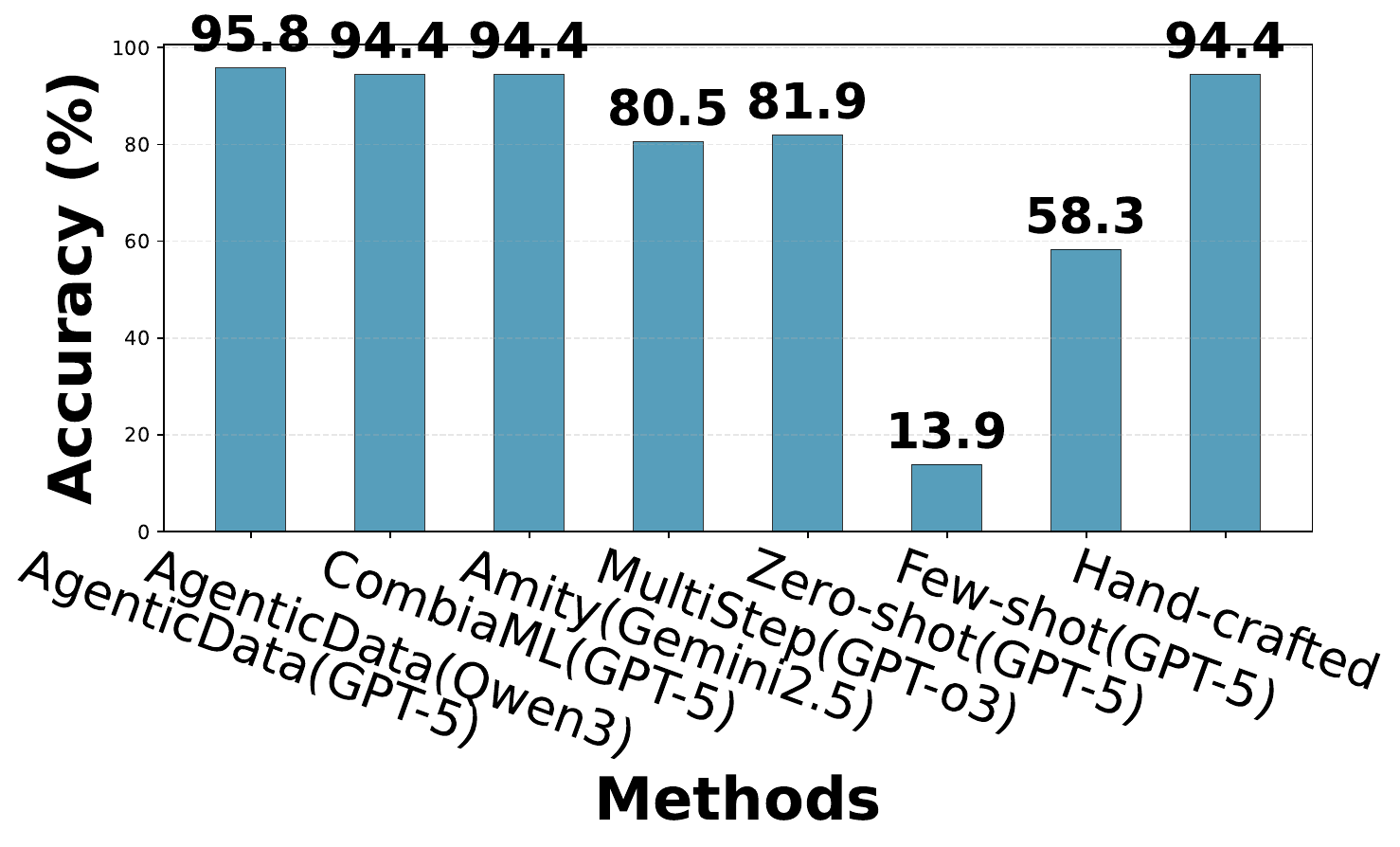}
        \label{fig:overall-dabstepeasy}
    }
    \subfigure[$\dabstep$-hard]{
        \includegraphics[width=0.15\linewidth]{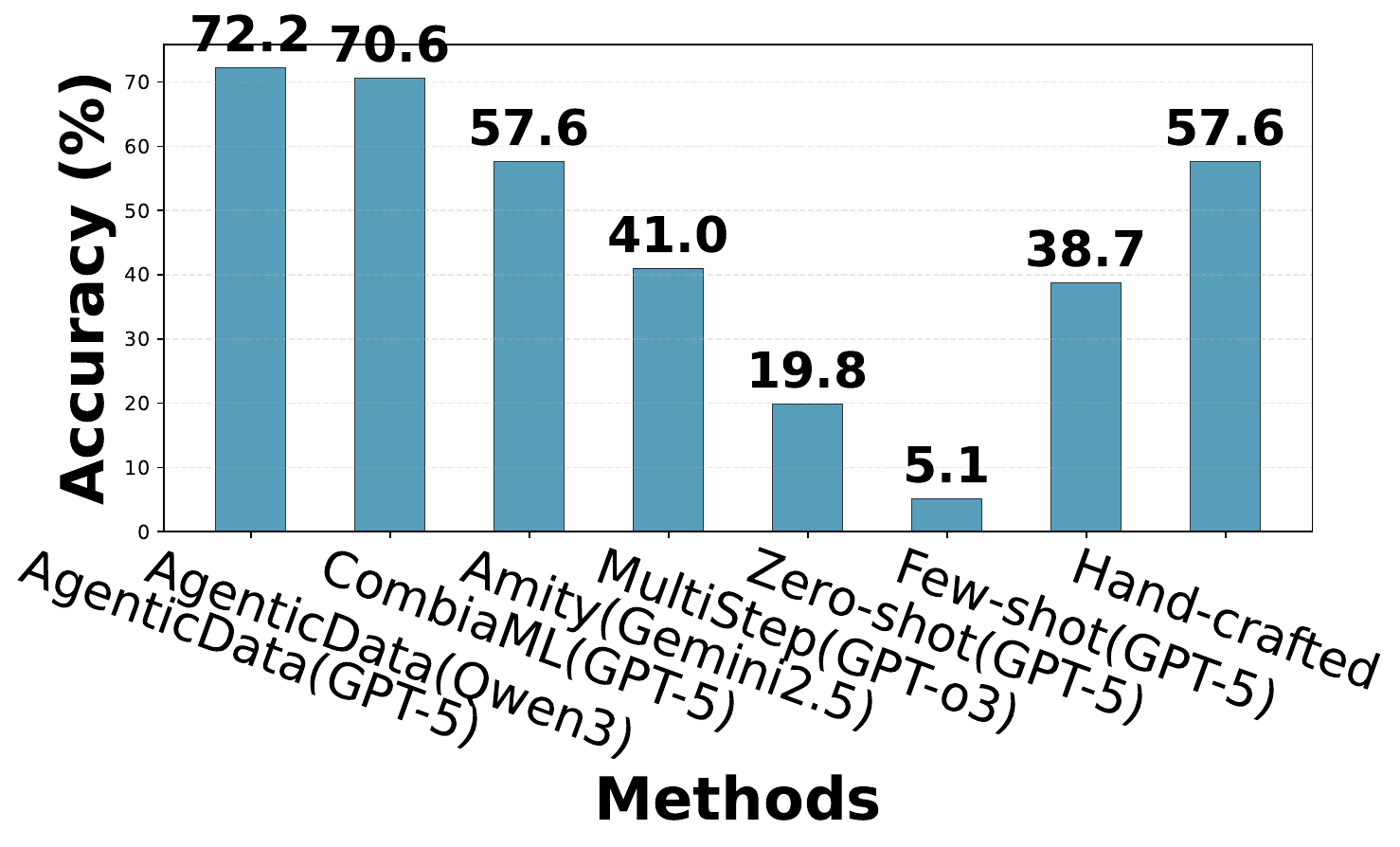}
        \label{fig:overall-dabstephard}
    }
    \subfigure[$\da$]{
        \includegraphics[width=0.15\linewidth]{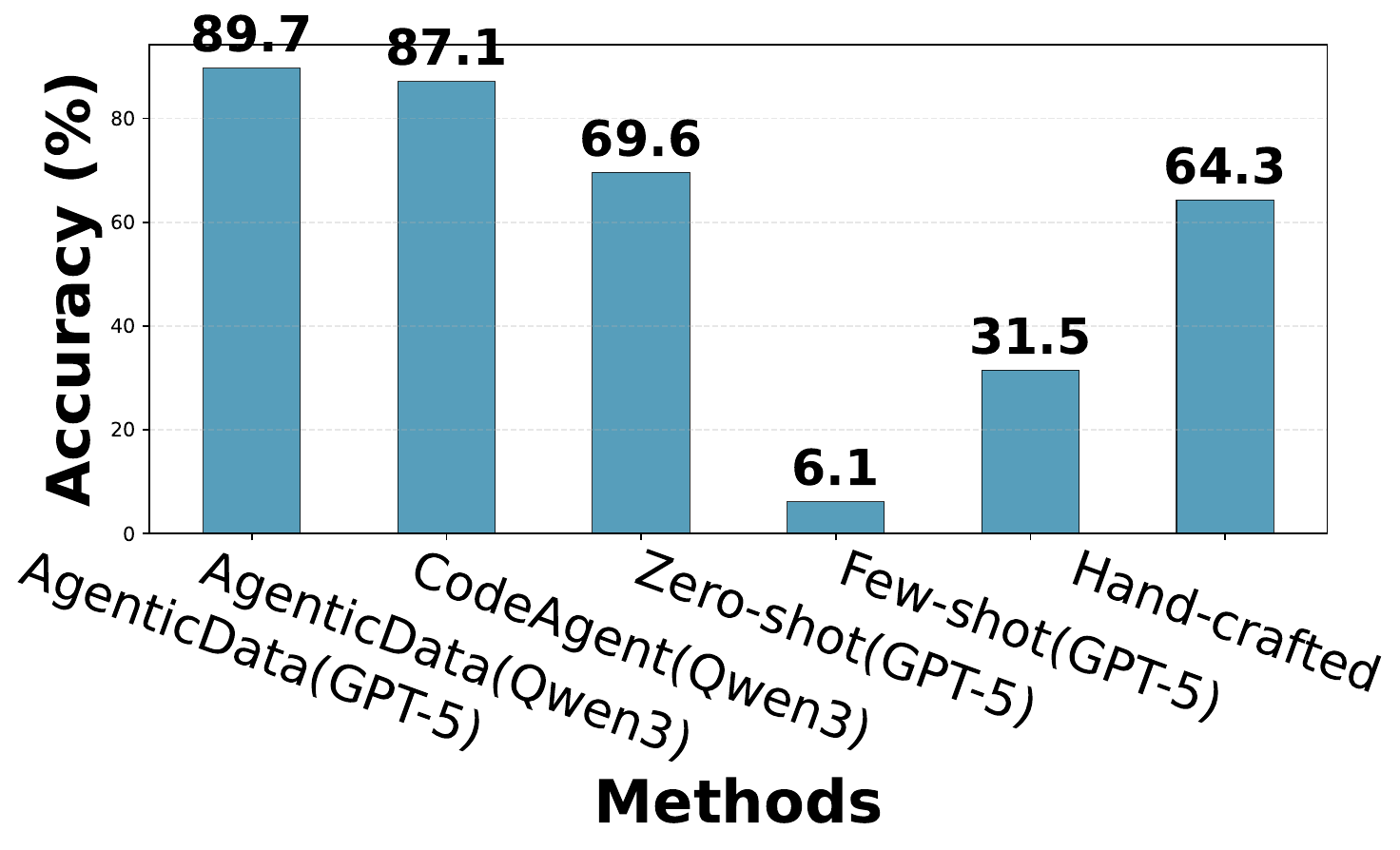}
        \label{fig:overall-dabench}
    }
    \subfigure[$\spider$]{
        \includegraphics[width=0.15\linewidth]{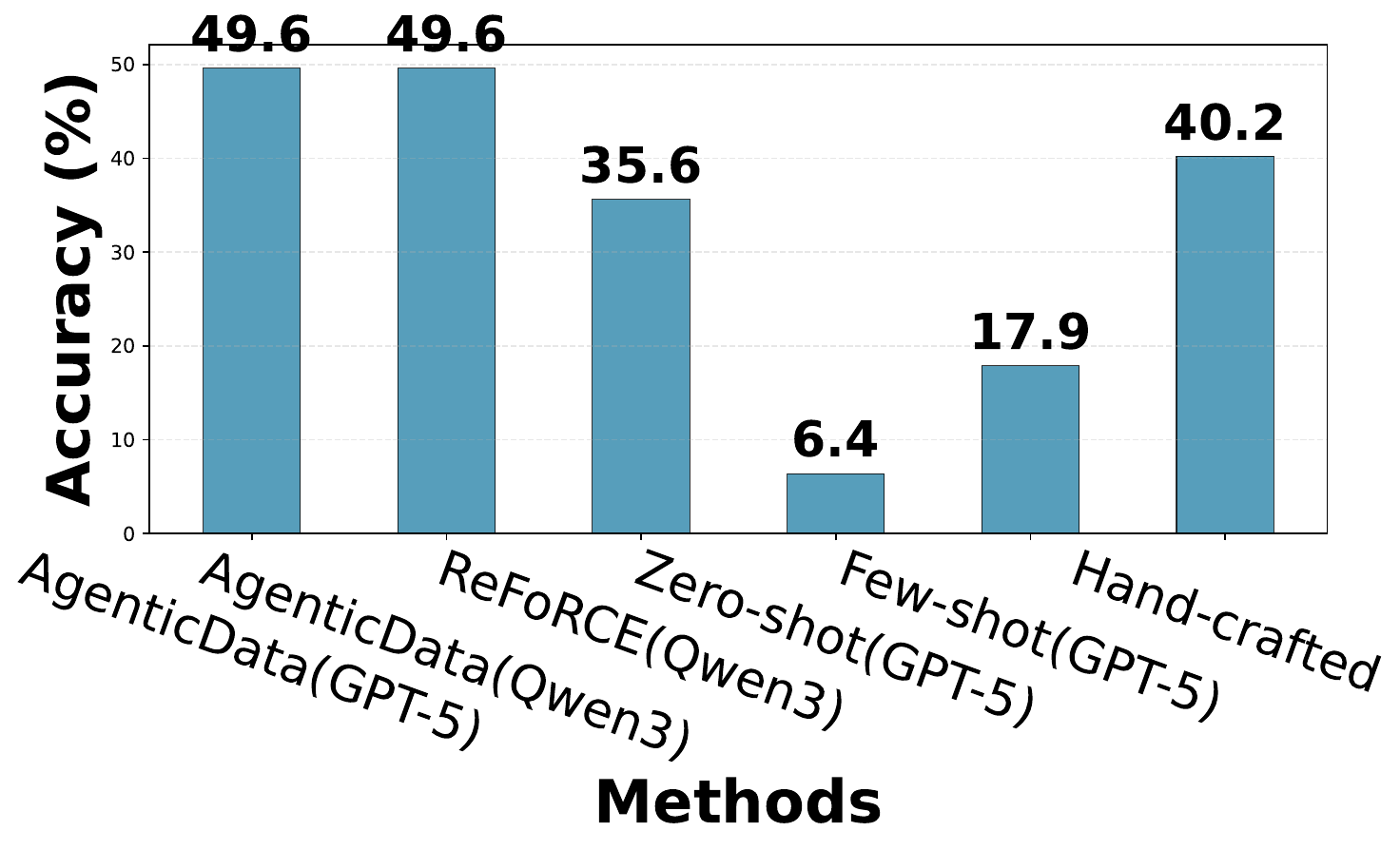}
        \label{fig:overall-spider}
    }
    \subfigure[$\bank$]{
        \includegraphics[width=0.15\linewidth]{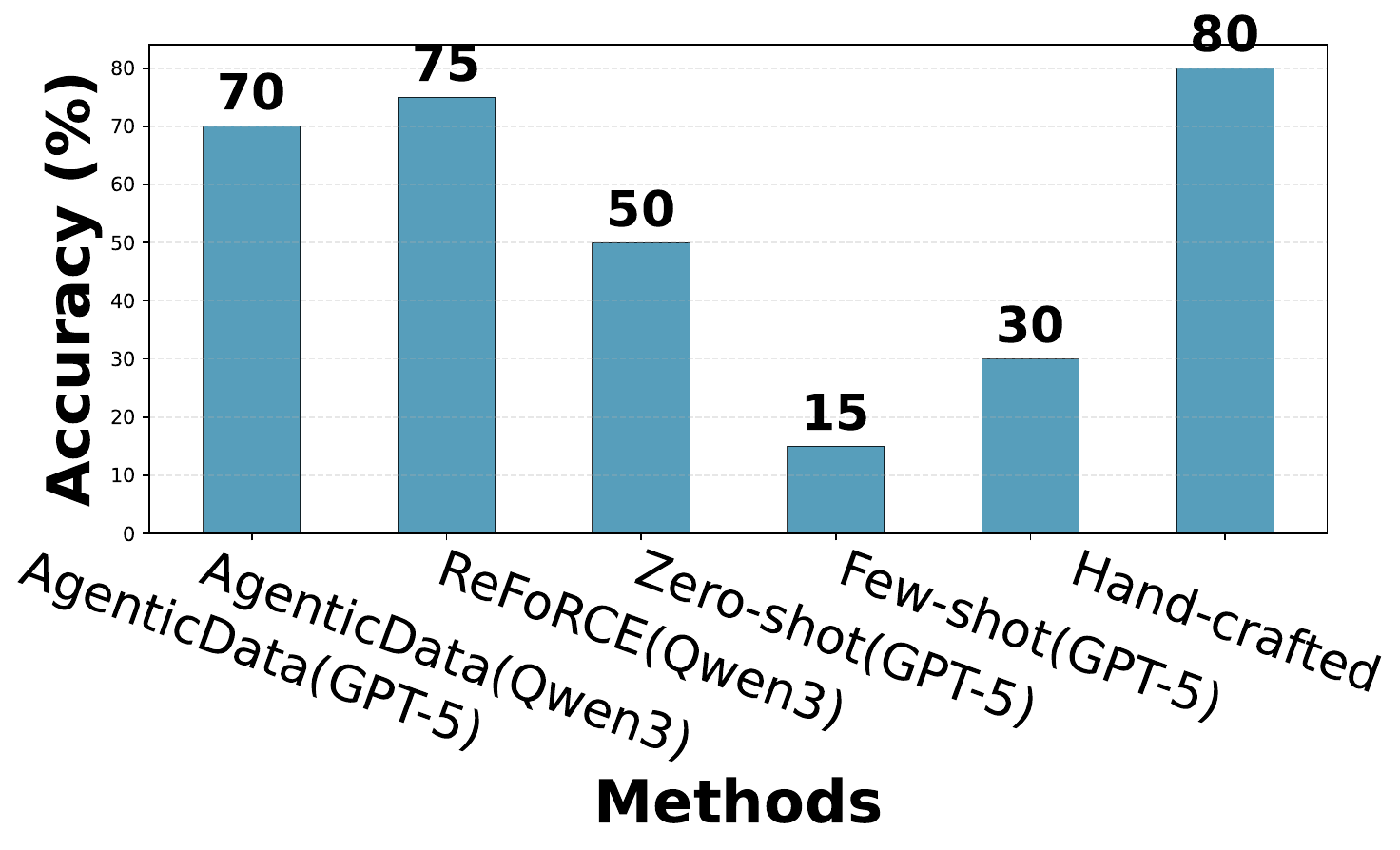}
        \label{fig:overall-bank}
    }
    \subfigure[$\wiki$]{
        \includegraphics[width=0.15\linewidth]{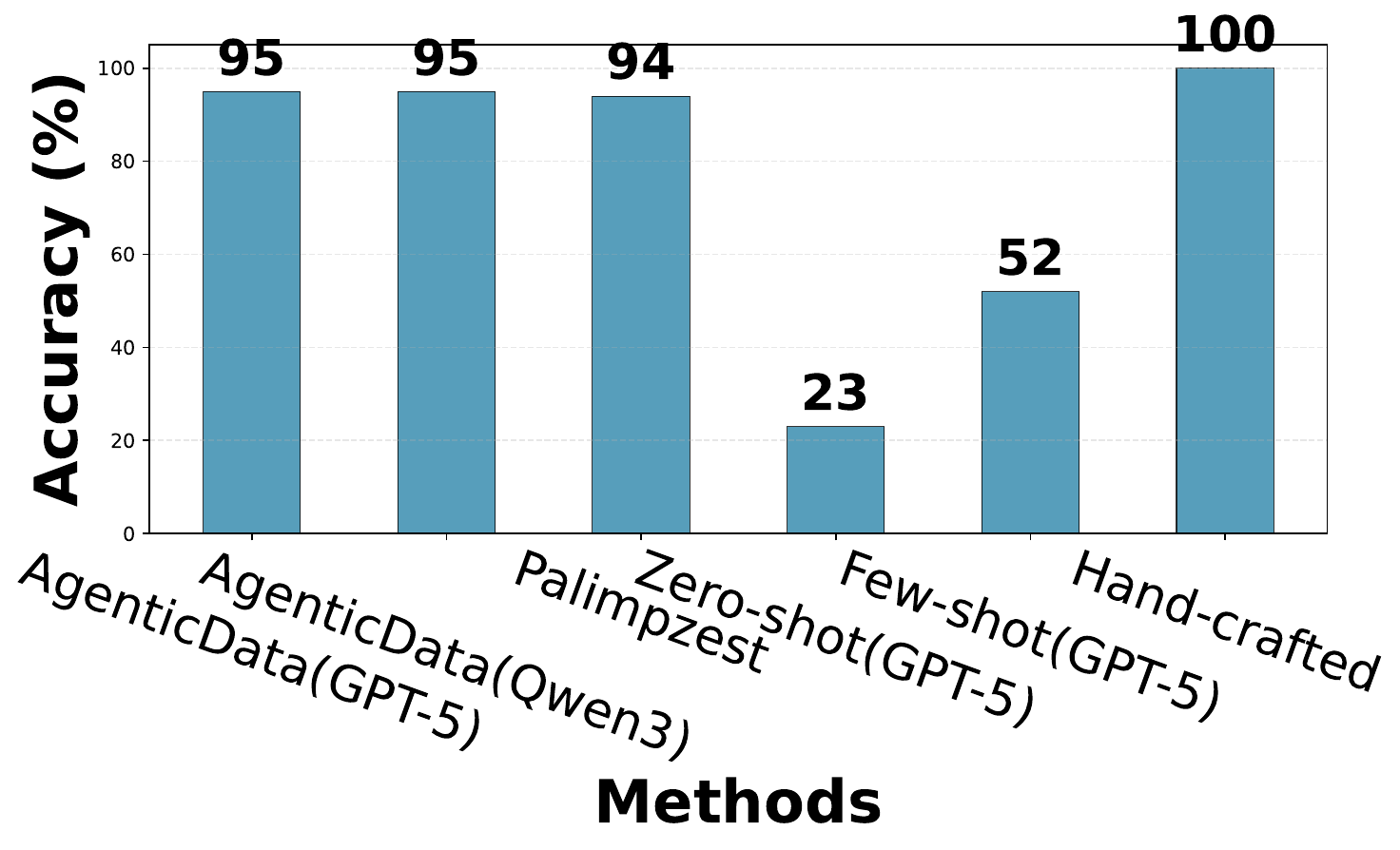}
        \label{fig:overall-wiki}
    }
    \vspace{-1.75em}
    \caption{Overall Accuracy Comparison on Benchmarks (\%).}\label{fig:overall}
        \vspace{-1.25em}
\end{figure*}

\begin{figure*}[!t]
    \centering
    \subfigure[$\dabstep$]{
    \includegraphics[width=0.18\linewidth]{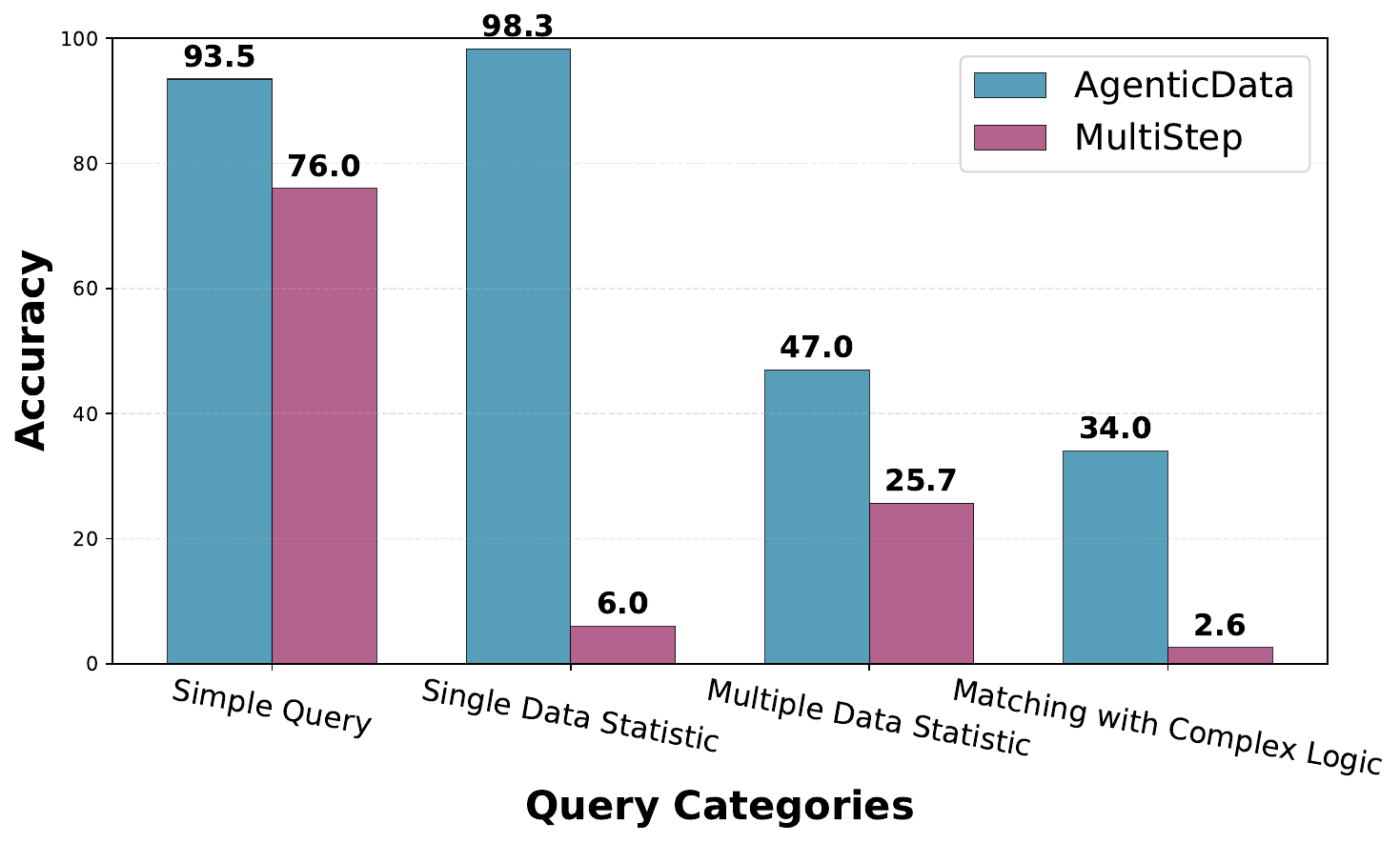}
    \label{fig:acc-dabstep}
    }
    \subfigure[$\da$]{
    \includegraphics[width=0.18\linewidth]{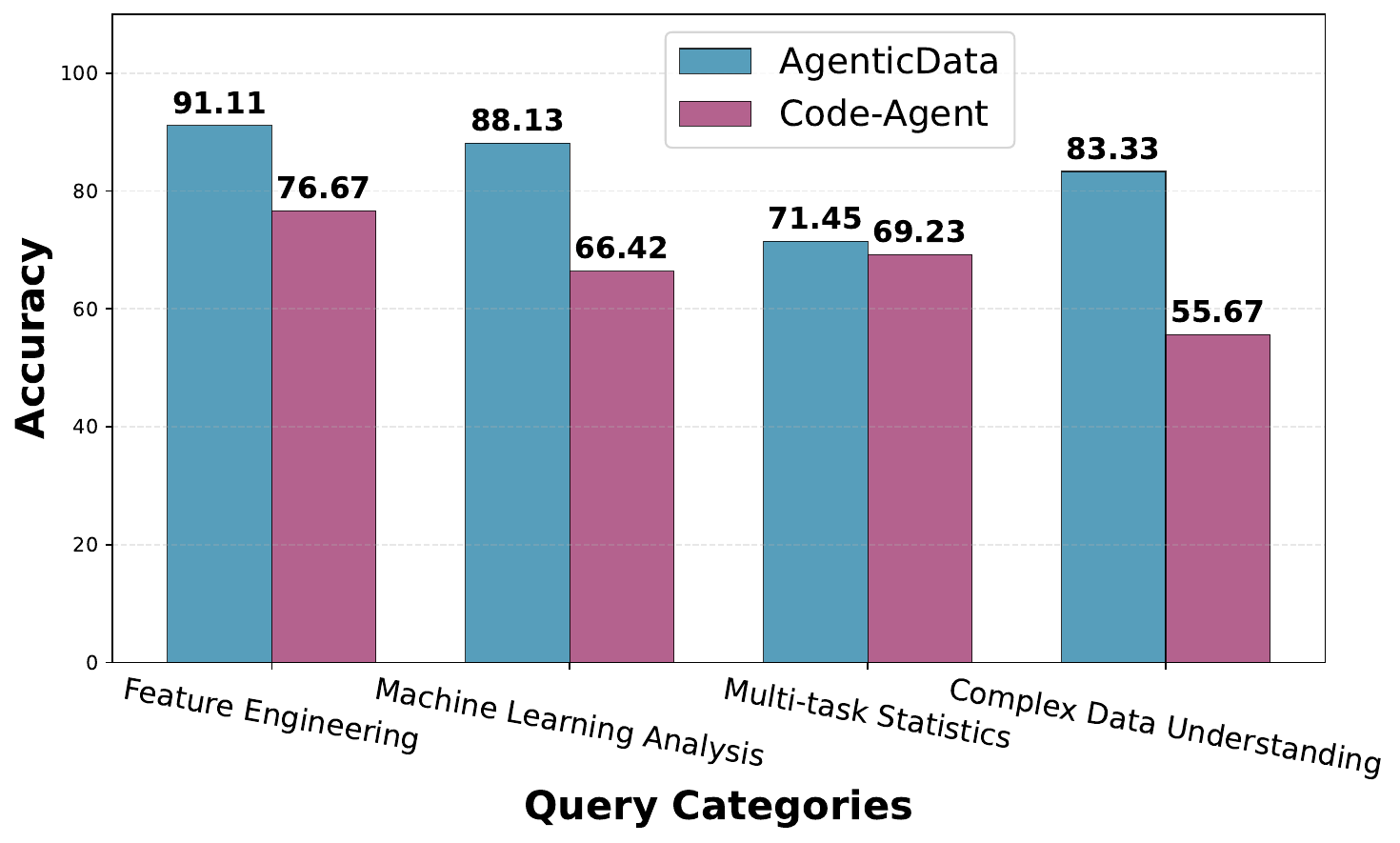}
    \label{fig:acc-da}
    }
    \subfigure[$\spider$]{
    \includegraphics[width=0.18\linewidth]{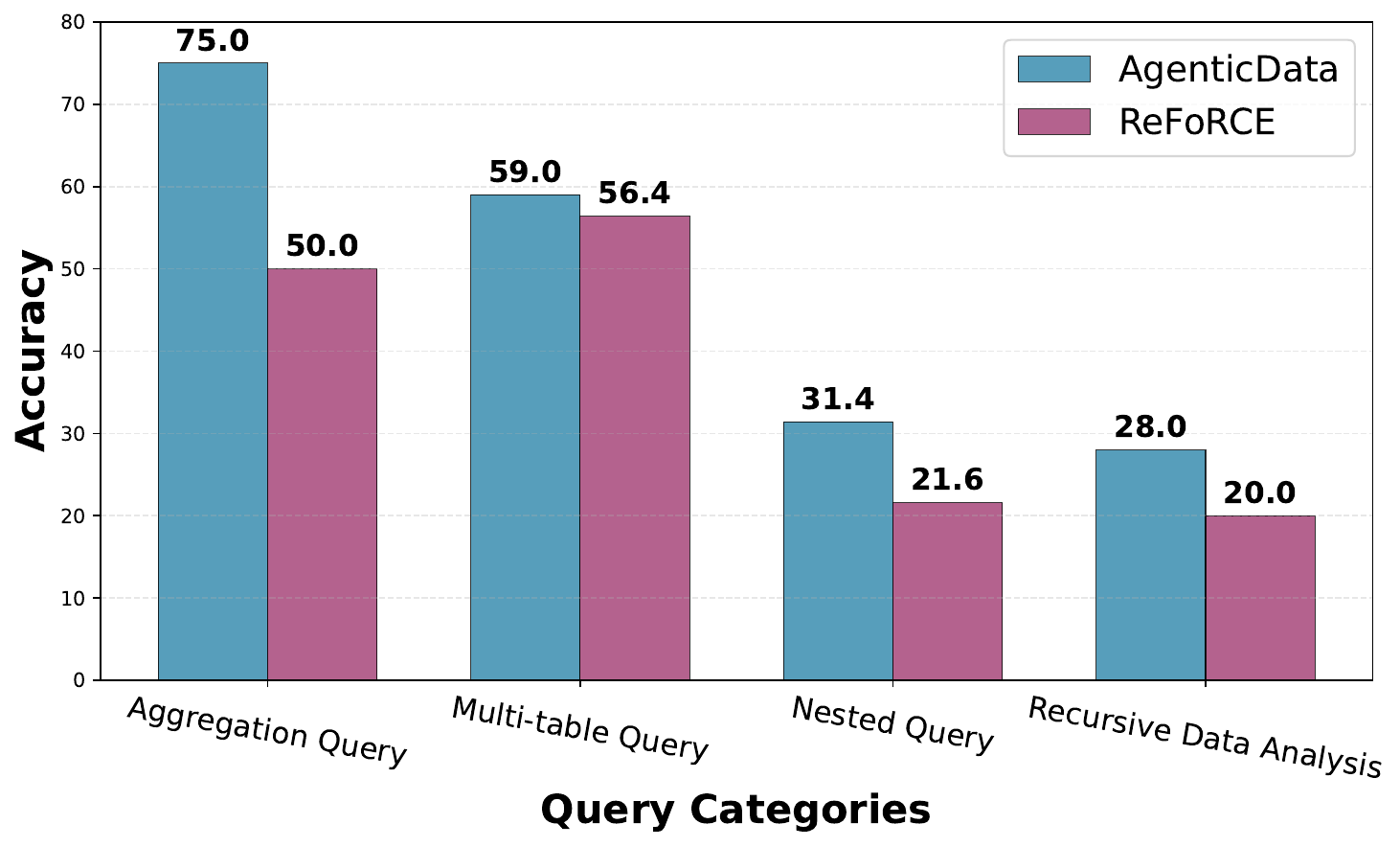}
    \label{fig:acc-spider}
    }
    \subfigure[$\bank$]{
    \includegraphics[width=0.18\linewidth]{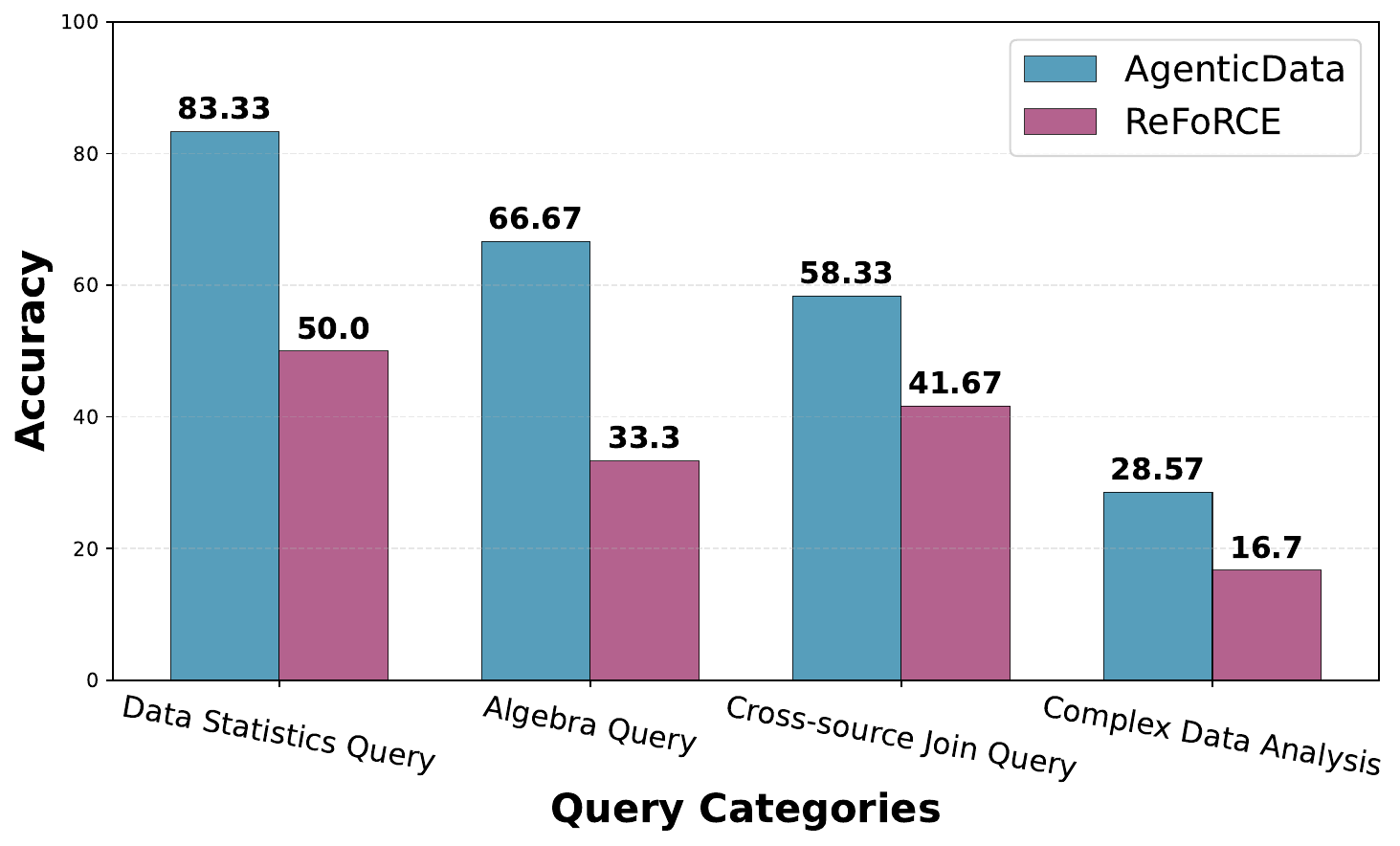}
    \label{fig:acc-bank}
   }
   \subfigure[$\wiki$]{
    \includegraphics[width=0.18\linewidth]{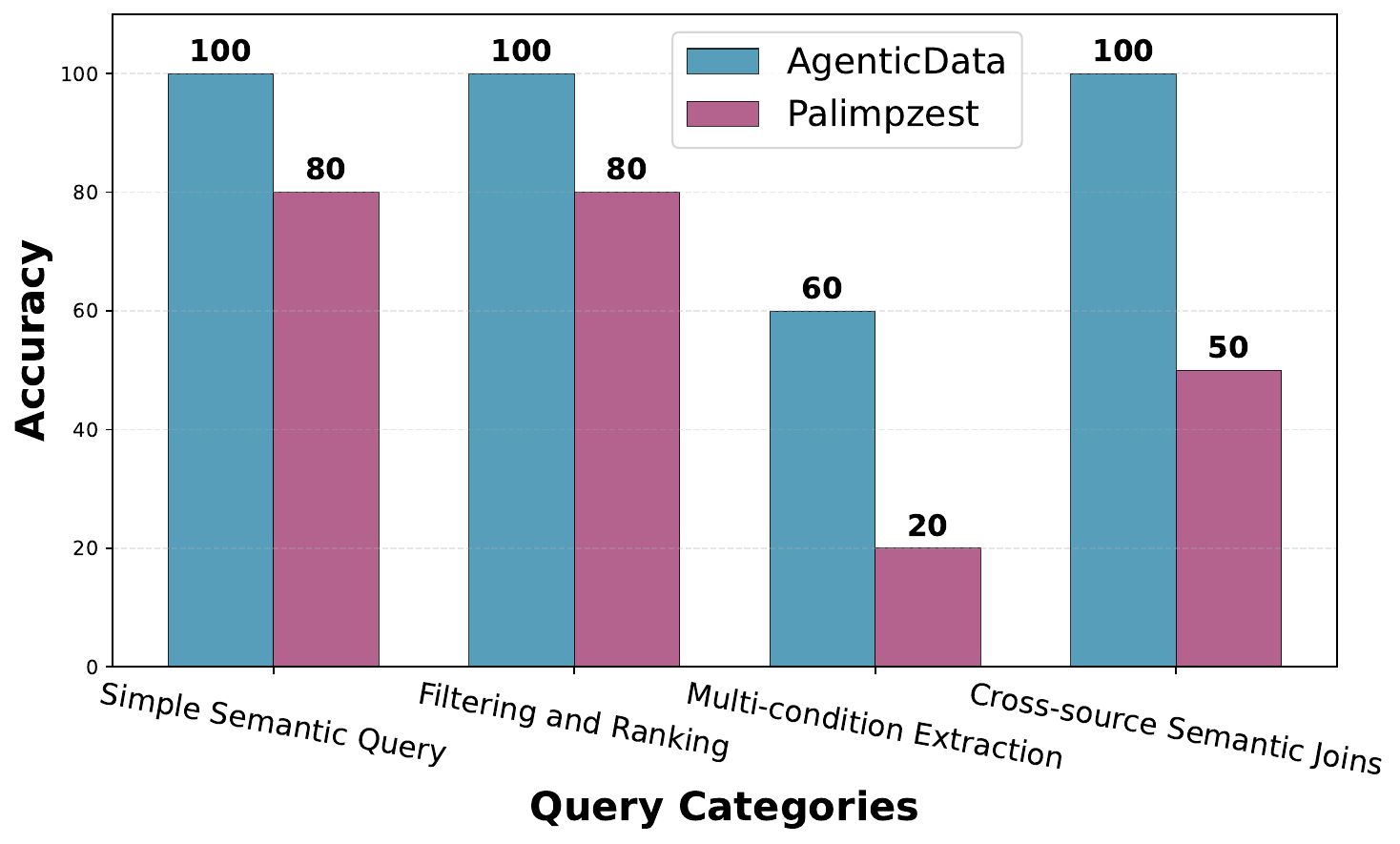}
    \label{fig:acc-wiki}
   }
   \vspace{-1.75em}
    \caption{Categorized Query Accuracy on Five Benchmarks. \label{fig:overall:accuracy}}
    \vspace{-1.25em}
\end{figure*}

\subsubsection{Overall Evaluation.}\label{sec:exp:overall}
We compared \our with state-of-the-art baselines. Figure~\ref{fig:overall} showed the overall accuracy. We made the following observations. First, \our outperformed its competitors by 13.8\% on easy tasks and 29.62\% on hard tasks on \dabstep. This advantage stemmed from \our's enhanced planning in understanding tasks and unstructured datasets, as well as its effective feedback through cross-validation and memory management. Notably, \our achieved significant accuracy improvements on difficult tasks on \dabstep by breaking down complex questions into sub-questions and constructing sub-plans to improve accuracy. 
Second, \our achieved a 17.51\% performance lead over \codeagent in advanced data analysis tasks on the \da benchmark. This was attributed to our semantic logical plan layer that links NL queries to code implementations, our effective validators to enhance accuracy, and our task decomposition techniques. 
Third, \our demonstrated a 14\% improvement over competitors in \spider, highlighting our support for multiple data sources and the capability of our feedback-based framework to formulate precise queries for complex data analysis on database engines. With \spider containing over 10,000 tables, the results confirmed that uncovering data linkages significantly enhances both data selection accuracy and overall precision. 
Fourth, in the case of real datasets and queries in \bank, \our outperformed \reforce by 19.75\% in accuracy. This improvement was due to its superior understanding of attributes within datasets with ambiguous column names through data sampling and terminology analysis, as well as by enhancing data profiles through mutual correlations between tables. 
Fifth, \our outperformed $\pz$~\cite{palimpzestCIDR} in unstructured data analytic tasks \wiki, because \our leveraged multi-agent techniques to effectively generate precise semantic query plans. Sixth, \zeroshot and \fewshot performed the worst on all the datasets because single-round prompting hardly generated accurate complex logic plans or operator parameters without experience memory. On benchmarks like \wiki and \bank, \manual had higher accuracy than other baselines, but it took too much human efforts comparing to agentic-based methods.
Finally, newer LLMs slightly improved the accuracy. While these newer LLMs performed better than older ones, they were still inferior to \our.

\vspace{-.5em}
\subsubsection{Evaluation on Different Types of Tasks.}

To evaluate the performance of \our across different types of tasks, we categorized these tasks and compared the accuracy for each category.

We classified \dabstep queries into four distinct categories, as depicted in Figure~\ref{fig:acc-dabstep}. \textit{Simple Queries} involved entire table statistics without including predicates or specific columns. \textit{Single Data Statistic Queries} required data statistics that involved intricate predicates and column alignment. \textit{Multiple Data Statistic Queries} were characterized by cross-data aggregation incorporating complex predicates and join conditions. \textit{Complex Logic Queries} necessitated the formulation of sophisticated rules to derive the targeted results. \our outperformed the baseline \smolagent across all task categories, notably in single data statistic and complex logic matching tasks. This success was attributed to the feedback-driven framework and sophisticated validation agents that adeptly identified semantic errors in predicates and refined logical plans. On some task types like complex matching queries, it was hard to write a right plan even for human, because there were many missing knowledge. For example, some tasks required data merging on multiple attributes, and there may be implicit knowledge for joins like "attribute ACI with NULL is a wildcard".

We categorized \da queries into four types, as shown in Figure~\ref{fig:acc-da}. \textit{Feature Engineering Queries} were responsible for extracting and transforming attributes, necessitating a thorough understanding of the task and precise selection of columns. \textit{Machine Learning Analysis Queries} relied on effective data preparation and the generation of machine learning code to interpret data. \textit{Multi-task Statistics Queries}  involved a sequence of tasks within one question, requiring advanced validation techniques to verify the accuracy of the task plan. \textit{Complex Data Understanding Queries} involved intricate filter conditions and aggregation functions across multiple attributes, necessitating highly accurate data profiling. \our significantly outperformed \codeagent on complex data analysis queries due to its intermediate logic semantic plan layer, which ensured the precision of code implementation.

We categorized queries in \spider into four groups, as shown in Figure~\ref{fig:acc-spider}. \textit{Aggregation Queries} focused on statistics from a single table and required an understanding of the task. \textit{Multi-table Queries}  involved combining data and computing statistics across several tables, necessitating careful data selection and identification of relationships between datasets. \textit{Nested Queries} included complex analytical tasks that require materialized views to clarify SQL logic. \textit{Recursive Data Analysis Queries} involved self-referential operations to navigate a graph-like data structure, demanding sophisticated logic validation skills. Our method outperformed baselines by 25\% on aggregation and recursive data analysis queries. This was due to our validation techniques for effectively rectifying intricate queries.

We categorized queries in \bank into four groups, as shown in Figure~\ref{fig:acc-bank}. \textit{Data Statistic Queries} comprised Year-on-Year and Month-on-Month statistics on loans data and required accurate data understanding. \textit{Algebra Queries} focused on algebra computations on financial data and required to profile columns precisely. \textit{Cross-source Join Queries} comprised analysis queries on multiple datasets and required thoroughly understanding the correlations between datasets. \textit{Complex Data Analysis Queries} included complex computation logic requiring sub-plan generation and sophisticated logic validation. Our method outperformed baselines by 100\% on algebra queries. This was because our data profiling techniques enhanced the accuracy of data selection and our planning method generated high-quality plans.

To evaluate the performance of various unstructured data analytic queries, we categorized queries on \wiki as illustrated in Figure~\ref{fig:acc-wiki}. \textit{Simple Semantic Queries} involved performing basic statistical summaries of the data and required a robust understanding of the questions along with the generation of semantic operators.  \textit{Filtering and Ranking Queries}  entailed selecting data based on specified conditions and ranking  according to the outcomes of given expressions, demanding advanced task understanding and logical validation capabilities. \textit{Multi-condition Extraction Queries} consisted of multiple filtering criteria for entity extraction, requiring precise generation of entity extraction operators. \textit{Cross-source Semantic Joins Queries} expanded unstructured data analysis to encompass multiple data sources, requiring the ability to identify related data across numerous datasets. \our outperformed the handcrafted \pz code by 200\% and 100\% in multi-condition extraction and cross-source semantic joins queries, because \our discovered related datasets, orchestrated high-quality plans, and generated superior semantic operators and logical plans.

\vspace{-.75em}
\subsubsection{Result Significance.}\label{sec:result-significance}
\begin{figure}[!t]\vspace{-.25em}
    \centering
    \includegraphics[width=0.9\linewidth]{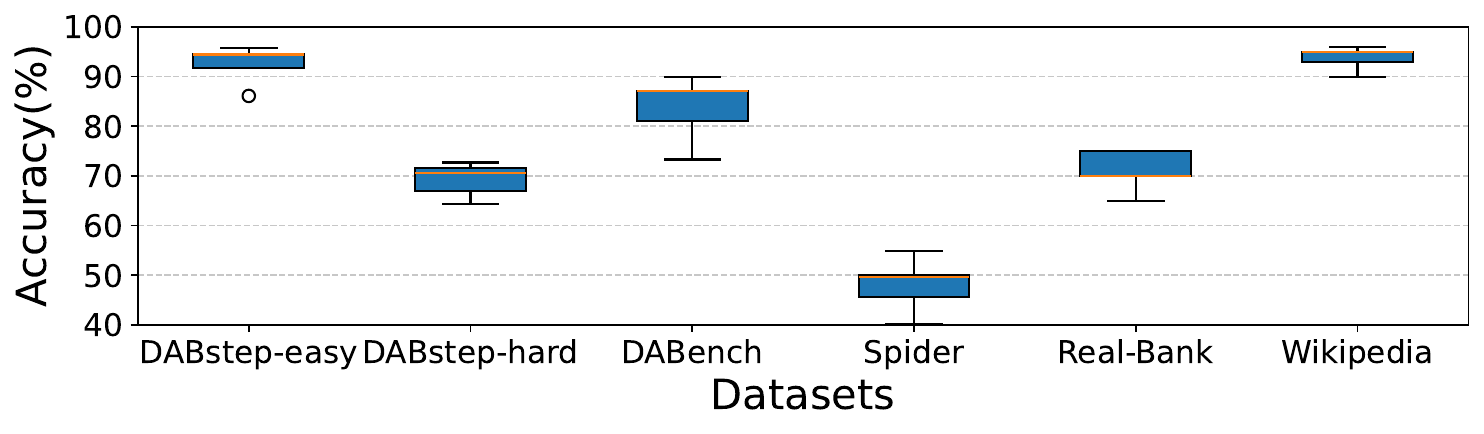}
        \vspace{-1.5em}
    \caption{Accuracy Distribution of \our}
    \label{fig:acc-variance}
    \vspace{-2em}
\end{figure}

Figure~\ref{fig:acc-variance} showed accuracy distribution across five repeated experiments conducted under identical configurations. The accuracy variance ranged from 5\% to 10\% in different benchmarks. Despite this variability, the median performance remained consistently high for all datasets.

\begin{figure*}[!t]\vspace{-2.5em}
    \centering
    \subfigure[$\dabstep$]{
        \includegraphics[width=0.185\linewidth]{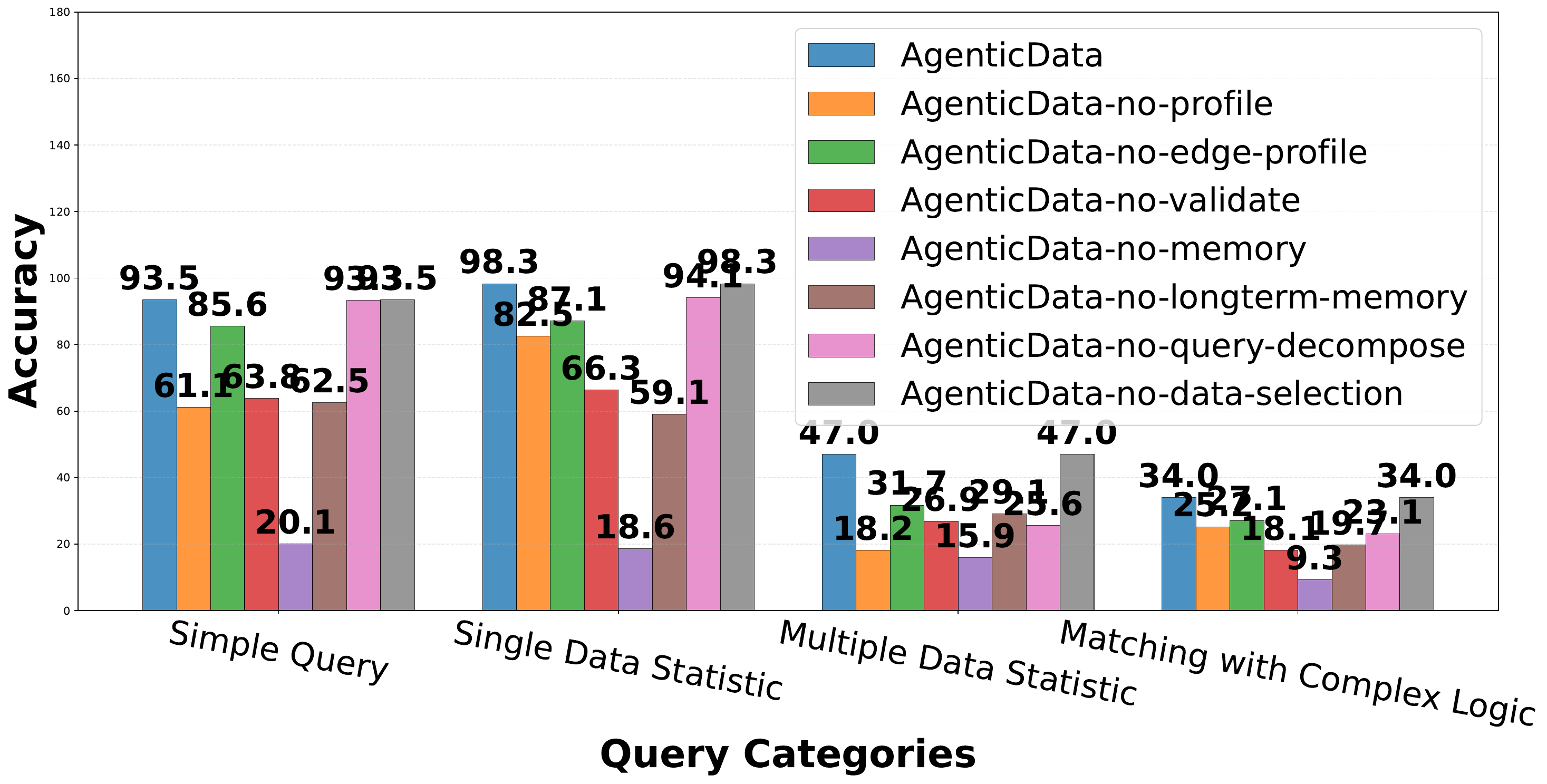}
        \label{fig:abl-dabstep}
    }
    \subfigure[$\da$]{
        \includegraphics[width=0.185\linewidth]{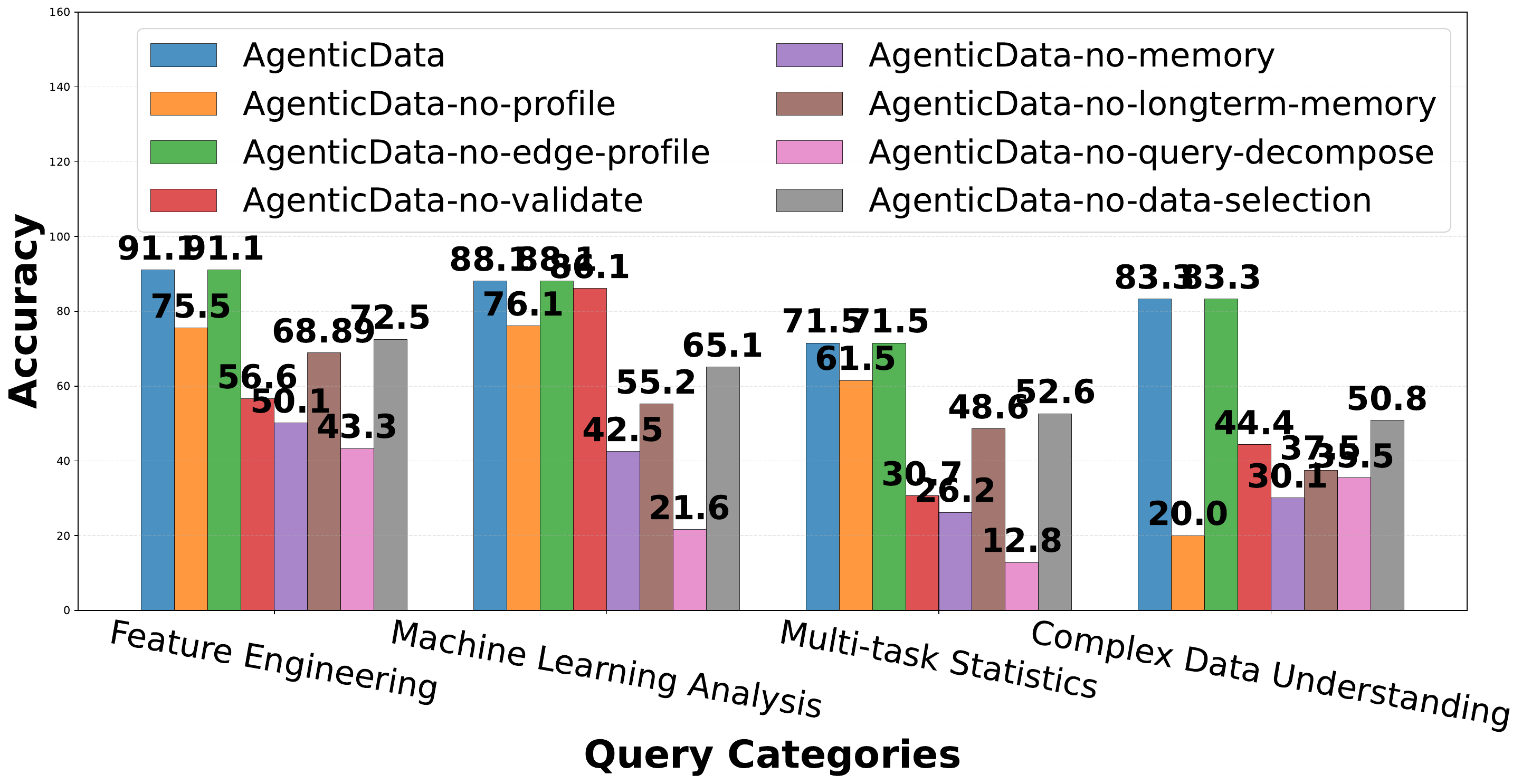}
        \label{fig:abl-dabench}
    }
    \subfigure[$\spider$]{
        \includegraphics[width=0.185\linewidth]{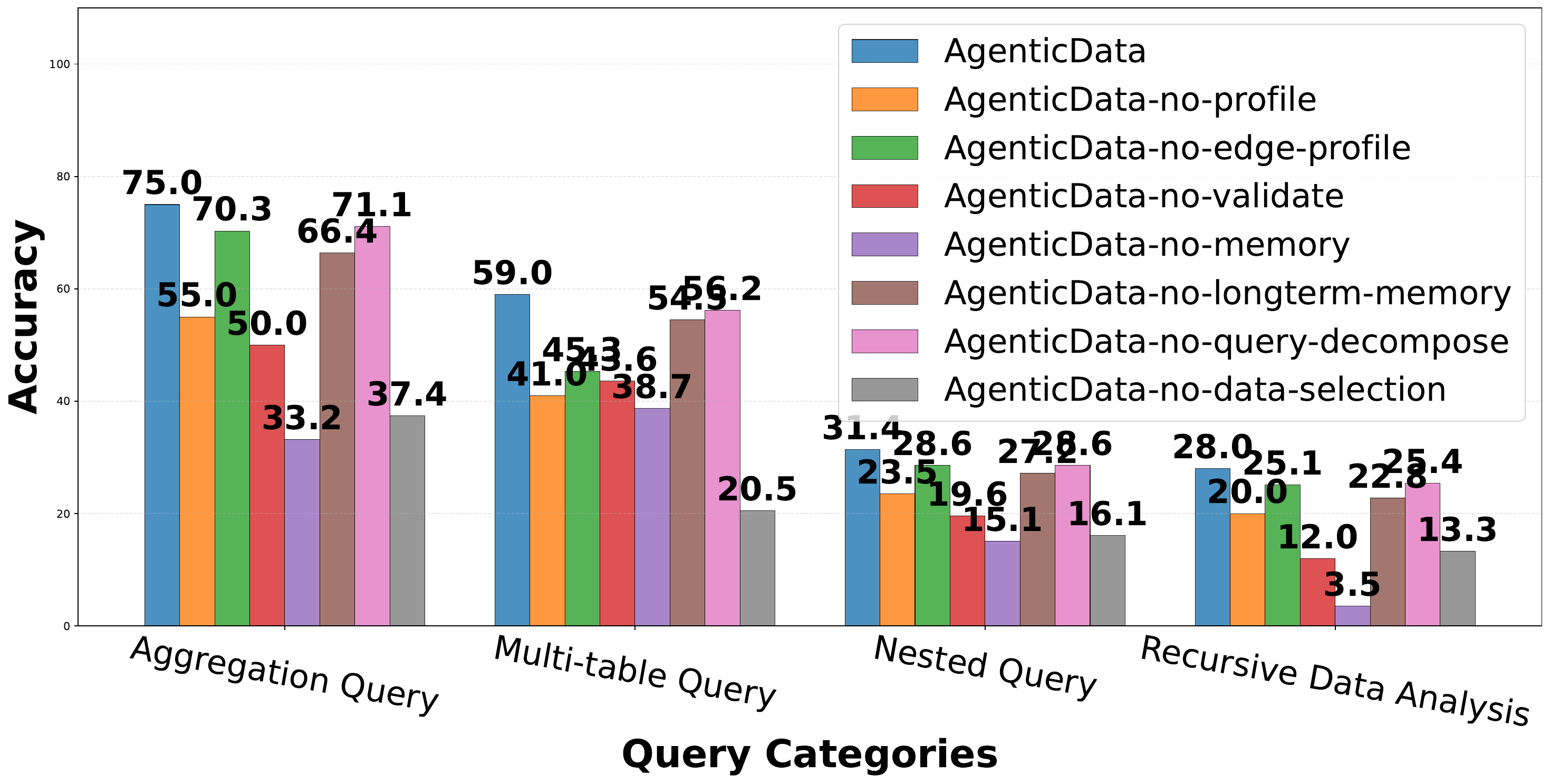}
        \label{fig:abl-spider}
    }
    \subfigure[$\bank$]{
        \includegraphics[width=0.185\linewidth]{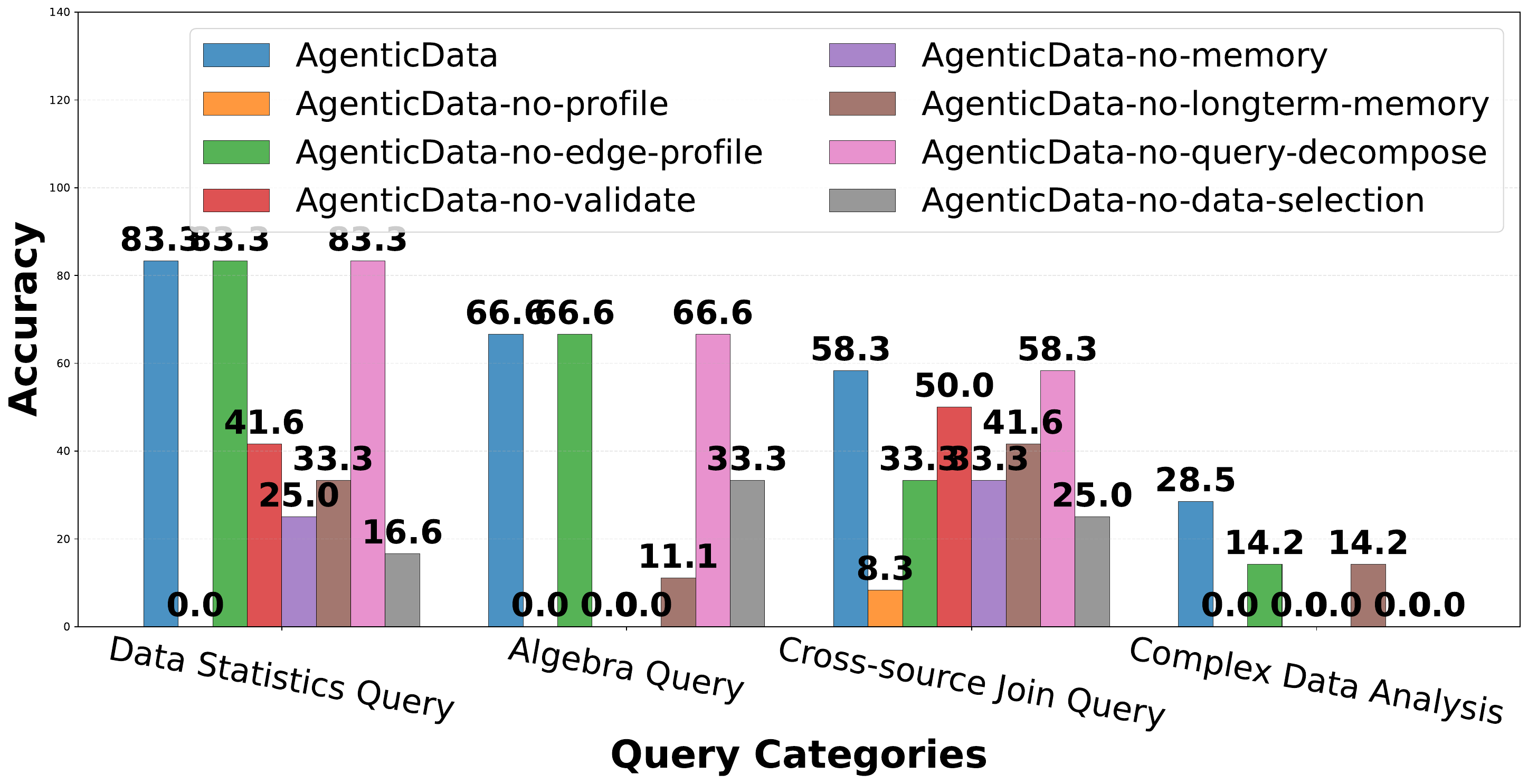}
        \label{fig:abl-bank}
    }
    \subfigure[$\wiki$]{
        \includegraphics[width=0.185\linewidth]{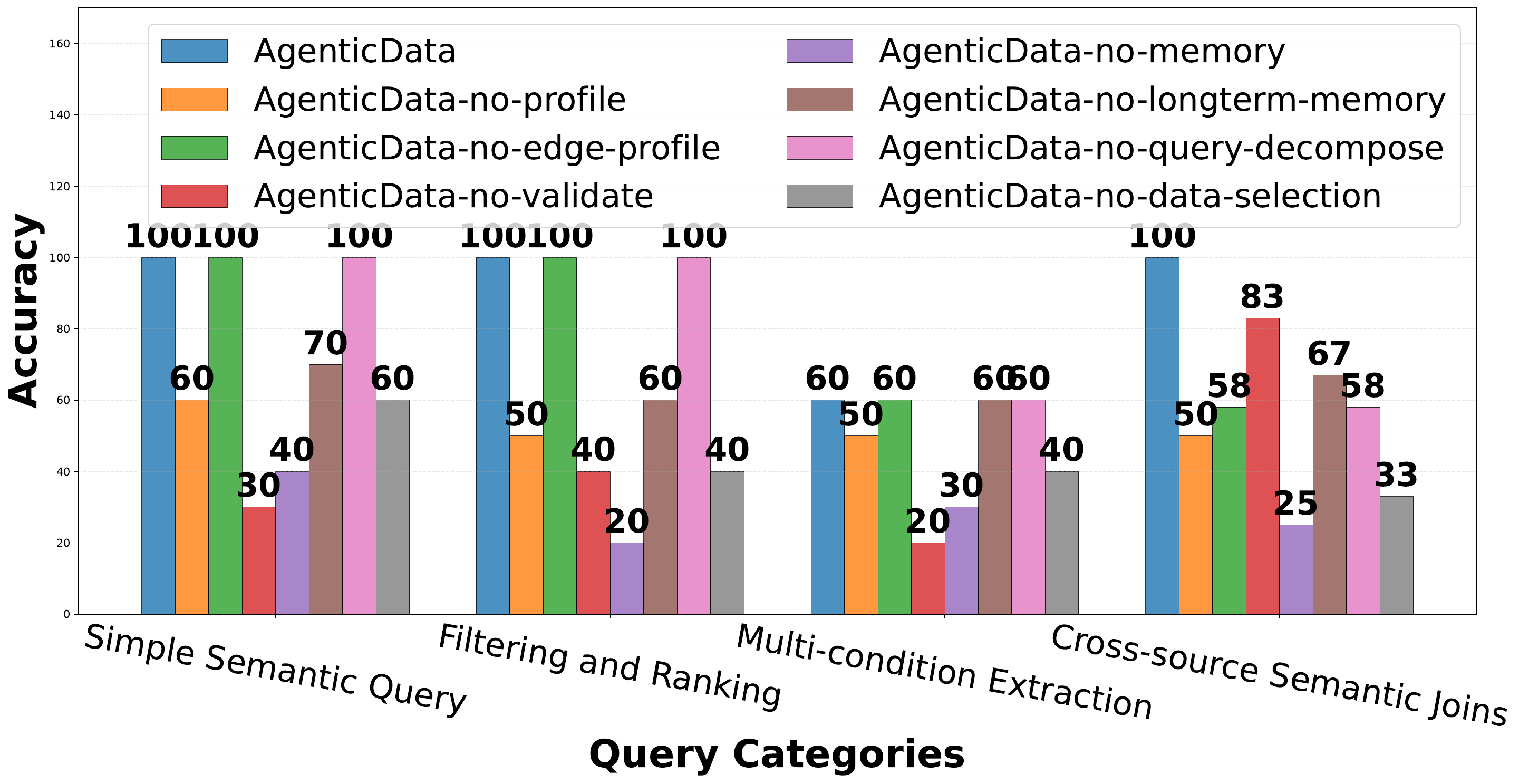}
        \label{fig:abl-wiki}
    }
    \vspace{-1.5em}
    \caption{Ablation Study On Different Tasks in Five Benchmarks.\label{fig:ablation}}
        \vspace{-1.5em}
\end{figure*}

\vspace{-.65em}
\subsection{Ablation Study on Different Tasks}\label{sec:exp:ablation}
\vspace{-.25em}


To assess the impact of our techniques on diverse task categories, we performed ablation studies on data profiling, plan validation, agentic memory, and query decomposition.

Figure~\ref{fig:abl-dabstep} illustrated the results on \dabstep. Agentic memory had the greatest impact on query accuracy because the system struggled to effectively correct errors based on feedback. Data selection had the least impact on accuracy, as the LLM could handle profiles of seven tables at once. Edge profiling improved accuracy by 10\%–20\% because some queries depended on attribute description files to enhance operator precision. Data profiling had a greater impact on accuracy than edge profiling because it also provided basic data insights for planning agents. The more complex the queries, the more query decomposition affected accuracy.




Figure~\ref{fig:abl-dabench} showed the results on \da. Query decomposition had the greatest impact, because these queries involved multiple attribute transformations and cannot be addressed with a single solution. Data profiling significantly influenced accuracy in complex data understanding queries by 80\%, because these queries required a precise understanding of data schema, semantics, and attribute types. Conversely, edge profiling had no significant impact on accuracy, as each task in \da utilized only one table. The agentic memory improved accuracy by 40\%–60\% across queries, as error correction was essential for handling complex tasks. Plan validation affected queries by identifying semantic errors. Long-term memory also influenced accuracy, as the summarized common knowledge helped reuse historical experience in the \da benchmark. Data selection was crucial to accuracy in \da because it could reduce the context size.

Figure~\ref{fig:abl-spider} showed the results on \spider. Data selection played a significant role in the accuracy of all queries by preventing context inflation. Our agentic memory had the greatest impact on improving accuracy across all query types, as \our relied on feedback using the memory mechanism to correct operators. However, long-term memory had little impact on accuracy because the queries involved various databases, making it difficult to share the memory effectively.



Figure~\ref{fig:abl-bank} showed the results from the real-world scenario \bank, which had two notable characteristics. First, a single table contained hundreds of columns. Second, the column names were often unclear, and data descriptions were typically insufficient. As a result, column matching posed the biggest challenge. When the data profiling agent was removed, the accuracy of data statistics queries, algebra queries, and complex data analysis queries dropped to zero, with overall accuracy falling to just 5\%. This was because understanding the data and selecting the appropriate columns became nearly impossible when relying only on superficial data schema information. Edge profiling improved the accuracy of cross-source join queries by 75.8\%, as precisely identifying correlated data was essential. Complex data analysis queries heavily depended on query decomposition to simplify plan generation. Long-term memory improved the accuracy of algebra queries by five times, as algebra expression generation benefited from shared experience.


Figure~\ref{fig:abl-wiki} showed that tasks within \wiki that was relied heavily on validation and agentic memory. Validation mechanisms improved the accuracy of simple semantic queries and filtering/ranking queries, because validators efficiently identify data and logic errors. 




\begin{figure}[!t]
    \centering
    \begin{minipage}[t]{0.23\textwidth}
        \centering
        \includegraphics[width=\textwidth]{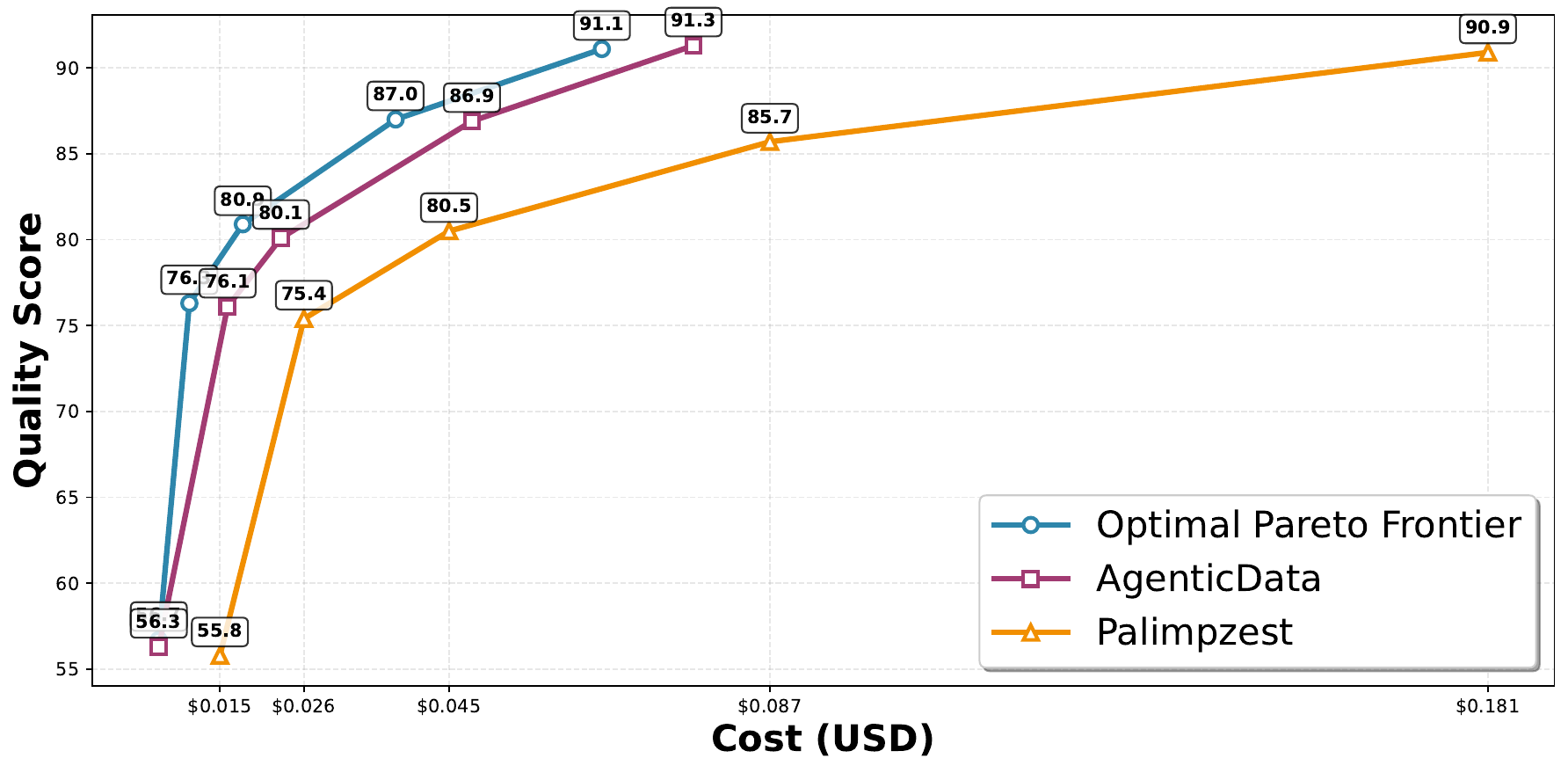}
        \vspace{-2em}
        \caption{Cost Comparison on \wiki.}\label{fig:cost-wiki}
    \end{minipage}
    \hfill 
    \begin{minipage}[t]{0.23\textwidth}
        \centering
        \includegraphics[width=\textwidth]{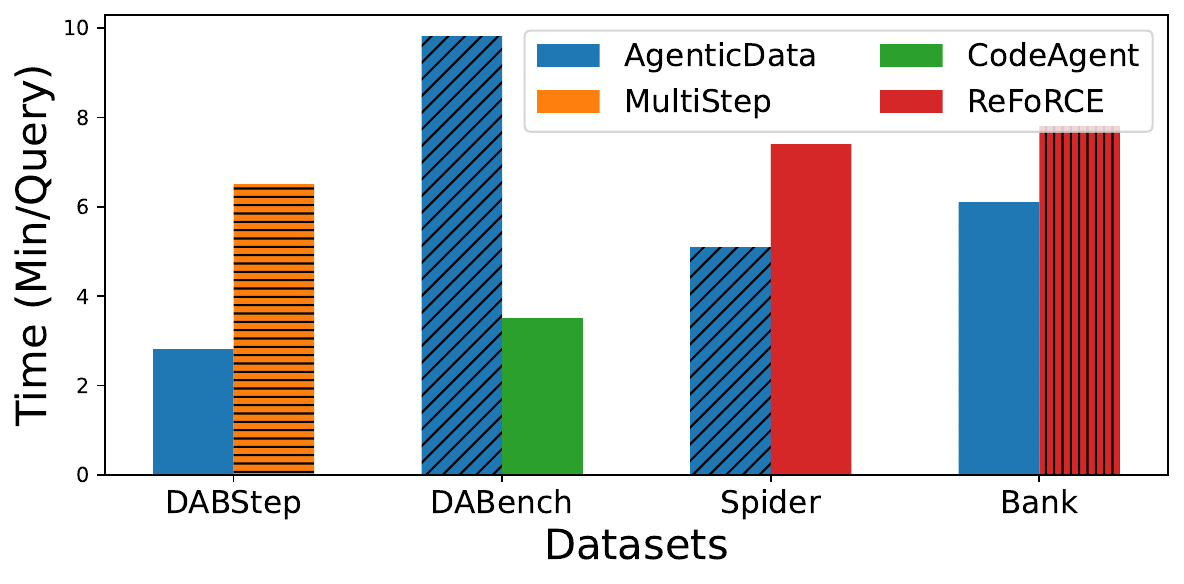}
        \vspace{-2em}
        \caption{Total Compute Time Comparison.}\label{fig:compare-time}
    \end{minipage}
    \vspace{-2em}
\end{figure}

%

%

\vspace{-.5em}
\subsection{Cost Comparison on Plan Optimization}\vspace{-.25em}

We compared quality and cost using the \wiki benchmark. We set quality thresholds at [55\%, 75\%, 80\%, 85\%, 90\%] and compared three methods to determine the minimum cost required to exceed each threshold. The Optimal Pareto Frontier represented the optimal plan of \our. As shown in Figure~\ref{fig:cost-wiki}, at each quality threshold, \our used only half or less of the cost compared to \pz and was close to the optimal cost. This improvement was due to our quality-aware cost optimization method.

\vspace{-.5em}
\subsection{Planning and Execution Latency}\label{sec:exp:latency}
\vspace{-.25em}

We evaluated planning and execution latency on the five benchmarks, and Table~\ref{tab:latency} showed the results. First, for planning time, existing methods took more than 10min, while \our achieved high planning  performance, taking 2.8min, 9.8min, 8.4min, 6.1min and 5.25min on benchmarks \dabstep, \da, \spider, \bank and \wiki respectively. This was attributed to our effective planning and feedback techniques. Second, for execution time,  \our achieved high execution performance, taking 11.74s, 29.34s, 0.42s, 0.0029s and 390.17s on benchmarks \dabstep, \da, \spider, \bank and \wiki respectively. These impressive results was attributed to our effective optimization and execution techniques. 

Figure~\ref{fig:compare-time} showed time comparison of \our and baselines. LLM inference time dominated the overall computation time. \our outperformed baselines on \dabstep, \spider and \bank, because our effective memory mechanism reduced the steps of planning. However, \codeagent was faster because it lacked precise data profiling process, but it had lower accuracy.

\subsection{Customer Real Use Cases}

We evaluated the performance of \our in real-world customer scenarios. At the State Grid Corporation of China, \our supported electrical load pressure analysis and marketing analytics on heterogeneous data  comprising hundreds of structured tables (with thousands columns) and tens of thousands of documents, achieving 91\% accuracy on over 200 questions, which was 30-40\% better than directly using LLMs and the ReAct agent framework. At the Postal Savings Bank of China, \our replaced the traditional BI system and answered ad-hoc complex analytical queries from all subsidiary managers. These queries involved complex data analysis and insights tasks on heterogeneous data, and \our achieved 92\% accuracy across 300 tasks, which was 35\% better than the ReAct agent framework.

\vspace{-.25em}
\section{Related Work}\label{sec:rw}
\vspace{-.25em}

\noindent\textbf{Agentic AI.} Reasoning LLMs~\cite{huang-chang-2023-towards,wei2023chainofthoughtpromptingelicitsreasoning,deepseekai2025deepseekr1incentivizingreasoningcapability,yao2023react} have gained significant attention, demonstrating remarkable abilities in handling intricate reasoning and planning tasks. 
Weng et al.~\cite{weng2023agent} devised a framework for an AI agent with four elements: planning, memory, tool utilization, and action. 
{\it However, these techniques achieved low accuracy on data analytics tasks because they could not  effectively generate the data profile and could not  support data analytics operators effectively.}

\begin{table}[!t]\vspace{-.35em}
    \centering
    \caption{Average Latency.}\vspace{-1.5em}
    \setlength{\tabcolsep}{1pt}
    {\footnotesize
    \begin{tabular}{|c|c|c|c|c|c|c|}
    \hline
        Benchmark & \dabstep & \da & Spider & \bank & \wiki \\\hline
        Planning  & 2.8min  & 9.8min & 5.1min  & 6.1min &  5.25min \\\hline
        Execution & 11.74s  & 29.34s  & 0.42s & 0.0029s & 390.17s \\\hline
    \end{tabular}
    \label{tab:latency}
    }
    \vspace{-1.5em}
\end{table}


\noindent\textbf{Semantic Data Analysis.}  Intelligent data analysis systems can be broadly categorized into four types: semantic query processing systems, NL2SQL systems, generic agent systems, and data agent systems. Lotus~\cite{Lotus} and Palimzest~\cite{palimpzestCIDR} are well-regarded for bridging the gap between structured and unstructured data via a unified operator framework; however, they still required skilled data analysts to script in declarative languages. Chat2DB utilized LLMs~\cite{chatdbglm} to generate SQL based on the task and  schema. Although NL2SQL systems can handle SQL dialects, they cannot perform cross-data-source analysis. Generic AI agent platforms like Smolagent~\cite{smolagents} and Xagent~\cite{xagent2023} used the ReAct framework to tackle complex tasks. {\it However, they achieved low accuracy on heterogeneous data because they focused on either unstructured or structured data but could not effectively generate the data profile for heterogeneous data.}


\noindent\textbf{Semantic Query Optimization. } To balance the cost and quality, researchers study the budget aware optimization problem for LLM tasks~\cite{shekhar2024optimizingcostsllmusage,russo2025abacuscostbasedoptimizersemantic}. Shekhar et al.~\cite{shekhar2024optimizingcostsllmusage} proposed a linear program method to route document segments to LLMs  for minimum cost or maximum quality. Russo et al.~\cite{russo2025abacuscostbasedoptimizersemantic} proposed a pareto-cascades to search for an optimal execution plan. {\it However, they faced challenges with a one-sided approach to quality estimation as they couldn't categorize the definition of high quality across different prompt classifications.}

\noindent\textbf{Memory Management.} Existing memory systems~\cite{zhang2024surveymemorymechanismlarge,chhikara2025mem0buildingproductionreadyai,xu2025amemagenticmemoryllm,kang2025memoryosaiagent} focused on the efficiency and accuracy of memory accessing. Mem0~\cite{chhikara2025mem0buildingproductionreadyai} supported large volumes of data, and A-Mem~\cite{xu2025amemagenticmemoryllm} dynamically organized memories to improve the system adaptation to different tasks. MemOS~\cite{kang2025memoryosaiagent} designed a hierarchical storage architecture for memory management. {\it However, these works could not assist LLMs in optimizing the state exploration reward and could not  provide evaluations for distinguishing between bad and good plans.}


\vspace{-.65em}
\section{Conclusions}
\vspace{-.125em}\vspace{-.25em}

We have proposed an agentic data analytics system, \our, which autonomously translates an NL query into a semantic query plan composed of semantic and relational operators. We have developed a multi-agent framework to generate high-quality semantic plans and introduced optimization techniques to improve the accuracy. Experimental results on widely-used benchmarks demonstrate that \our significantly outperformed state-of-the-art baselines. \our has been deployed in real data analysis applications within banks and power grid companies.


 \bibliographystyle{abbrv}

\newpage
\balance
\bibliography{acmart}

@article{DocETL,
      title={DocETL: Agentic Query Rewriting and Evaluation for Complex Document Processing}, 
      author={Shreya Shankar and Tristan Chambers and Tarak Shah and Aditya G. Parameswaran and Eugene Wu},
      year={2025},
      journal={VLDB},
      eprint={2410.12189},
      archivePrefix={arXiv},
      primaryClass={cs.DB},
      url={https://arxiv.org/abs/2410.12189}, 
}

@article{kang2025memoryosaiagent,
      title={Memory OS of AI Agent}, 
      author={Jiazheng Kang and Mingming Ji and Zhe Zhao and Ting Bai},
      year={2025},
      eprint={2506.06326},
      journal={arXiv},
      primaryClass={cs.AI},
      url={https://arxiv.org/abs/2506.06326}, 
}

@article{DBLP:journals/pvldb/UrbanB24,
  author       = {Matthias Urban and
                  Carsten Binnig},
  title        = {{ELEET:} Efficient Learned Query Execution over Text and Tables},
  journal      = {Proc. {VLDB} Endow.},
  volume       = {17},
  number       = {13},
  pages        = {4867--4880},
  year         = {2024},
  url          = {https://www.vldb.org/pvldb/vol17/p4867-urban.pdf},
  bibsource    = {dblp computer science bibliography, https://dblp.org}
}

@article{DBLP:journals/corr/abs-2506-06541,
  author       = {Eugenie Lai and
                  Gerardo Vitagliano and
                  Ziyu Zhang and
                  Sivaprasad Sudhir and
                  Om Chabra and
                  Anna Zeng and
                  Anton A. Zabreyko and
                  Chenning Li and
                  Ferdi Kossmann and
                  Jialin Ding and
                  Jun Chen and
                  Markos Markakis and
                  Matthew Russo and
                  Weiyang Wang and
                  Ziniu Wu and
                  Michael J. Cafarella and
                  Lei Cao and
                  Samuel Madden and
                  Tim Kraska},
  title        = {KramaBench: {A} Benchmark for {AI} Systems on Data-to-Insight Pipelines
                  over Data Lakes},
  journal      = {CoRR},
  volume       = {abs/2506.06541},
  year         = {2025},
  url          = {https://doi.org/10.48550/arXiv.2506.06541},
  doi          = {10.48550/ARXIV.2506.06541},
  eprinttype    = {arXiv},
  eprint       = {2506.06541},
  bibsource    = {dblp computer science bibliography, https://dblp.org}
}

@inproceedings{DBLP:conf/cidr/UrbanB24,
  author       = {Matthias Urban and
                  Carsten Binnig},
  title        = {{CAESURA:} Language Models as Multi-Modal Query Planners},
  booktitle    = {CIDR},
  publisher    = {www.cidrdb.org},
  year         = {2024},
  url          = {https://www.cidrdb.org/cidr2024/papers/p14-urban.pdf},
  bibsource    = {dblp computer science bibliography, https://dblp.org}
}

@misc{RealBank, 
    title = {Real-Bank},
    note={https://anonymous.4open.science/r/AgenticData-696B}
}

@misc{Spider2,
      title = {Spider 2},
      note={https://spider2-sql.github.io}}

@misc{DABench,
      title = {DABench},
      note={https://infiagent.github.io/}}

@misc{WIKI,
      title = {Wiki},
      note={https://dumps.wikimedia.org/}}

@misc{DABStepURL,
      title = {DABStep},
      note={https://huggingface.co/datasets/adyen/DABstep}}

@article{Lotus,
      title={Semantic Operators: A Declarative Model for Rich, AI-based Analytics Over Text Data},
      author={Liana Patel and Siddharth Jha and Parth Asawa and Melissa Pan and Carlos Guestrin and Matei Zaharia},
      year={2024},
      eprint={2407.11418},
      journal={arXiv},
      primaryClass={cs.DB},
      url={https://arxiv.org/abs/2407.11418},
}

@article{Luna,
  title={The Design of an LLM-powered Unstructured Analytics System},
  author={Anderson, Eric and Fritz, Jonathan and Lee, Austin and Li, Bohou and Lindblad, Mark and Lindeman, Henry and Meyer, Alex and Parmar, Parth and Ranade, Tanvi and Shah, Mehul A and others},
  journal={arXiv preprint arXiv:2409.00847},
  year={2024}
}

@inproceedings{ZenDB,
      title={Towards Accurate and Efficient Document Analytics with Large Language Models}, 
      author={Yiming Lin and Madelon Hulsebos and Ruiying Ma and Shreya Shankar and Sepanta Zeigham and Aditya G. Parameswaran and Eugene Wu},
      year={2025},
      booktitle={ICDE},
}

@inproceedings{TAG,
  title={Text2sql is not enough: Unifying ai and databases with tag},
  author={Biswal, Asim and Patel, Liana and Jha, Siddarth and Kamsetty, Amog and Liu, Shu and Gonzalez, Joseph E and Guestrin, Carlos and Zaharia, Matei},
  booktitle={CIDR},
  year={2025}
}

@inproceedings{palimpzestCIDR,
    title={Palimpzest: Optimizing AI-Powered Analytics with Declarative Query Processing},
    author={Liu, Chunwei and Russo, Matthew and Cafarella, Michael and Cao, Lei and Chen, Peter Baile and Chen, Zui and Franklin, Michael and Kraska, Tim and Madden, Samuel and Shahout, Rana and Vitagliano, Gerardo},
    booktitle = {CIDR},
    year = 2025,
}

@inproceedings{UQE,
      title={UQE: A Query Engine for Unstructured Databases}, 
      author={Hanjun Dai and Bethany Yixin Wang and Xingchen Wan and Bo Dai and Sherry Yang and Azade Nova and Pengcheng Yin and Phitchaya Mangpo Phothilimthana and Charles Sutton and Dale Schuurmans},
      year={2025},
     booktitle = {CIDR},
}

@inproceedings{AOP,
  title={Aop: Automated and interactive llm pipeline orchestration for answering complex queries},
  author={Wang, Jiayi and Li, Guoliang},
  year={2025},
  booktitle={CIDR}
}

@INPROCEEDINGS{unify,
  author={Wang, Jiayi and Li, Guoliang and Feng, Jianhua},
  booktitle={42th International Conference on Data Engineering (ICDE)}, 
  title={Unify: An Unstructured Data Analytics System}, 
  year={2025},
  volume={},
  number={},
  pages={},
}

@article{liu2025surveytexttosqlerallms,
      title={A Survey of Text-to-SQL in the Era of LLMs: Where are we, and where are we going?}, 
      author={Xinyu Liu and Shuyu Shen and Boyan Li and Peixian Ma and Runzhi Jiang and Yuxin Zhang and Ju Fan and Guoliang Li and Nan Tang and Yuyu Luo},
      year={2025},
      journal={IEEE TKDE}
}

@article{10.14778/3685800.3685905,
author = {Zhao, Xinyang and Zhou, Xuanhe and Li, Guoliang},
title = {Chat2Data: An Interactive Data Analysis System with RAG, Vector Databases and LLMs},
year = {2024},
issue_date = {August 2024},
publisher = {VLDB},
volume = {17},
number = {12},
issn = {2150-8097},
url = {https://doi.org/10.14778/3685800.3685905},
doi = {10.14778/3685800.3685905},
journal = {Proc. VLDB Endow.},
month = aug,
pages = {4481–4484},
numpages = {4}
}

@inproceedings{huang-chang-2023-towards,
    title = "Towards Reasoning in Large Language Models: A Survey",
    author = "Huang, Jie  and
      Chang, Kevin Chen-Chuan",
    booktitle = "ACL",
    year = "2023",
    url = "https://aclanthology.org/2023.findings-acl.67/",
    doi = "10.18653/v1/2023.findings-acl.67",
    pages = "1049--1065"
}

@article{wei2023chainofthoughtpromptingelicitsreasoning,
      title={Chain-of-Thought Prompting Elicits Reasoning in Large Language Models}, 
      author={Jason Wei and Xuezhi Wang and Dale Schuurmans and Maarten Bosma and Brian Ichter and Fei Xia and Ed Chi and Quoc Le and Denny Zhou},
      year={2023},
      eprint={2201.11903},
      journal={arXiv},
      primaryClass={cs.CL},
      url={https://arxiv.org/abs/2201.11903}, 
}

@article{deepseekai2025deepseekr1incentivizingreasoningcapability,
      title={DeepSeek-R1: Incentivizing Reasoning Capability in LLMs via Reinforcement Learning}, 
      author={DeepSeek-AI and Daya Guo et al},
      year={2025},
      eprint={2501.12948},
      journal={arXiv},
      primaryClass={cs.CL},
      url={https://arxiv.org/abs/2501.12948}, 
}

@inproceedings{yao2023react,
  title = {{ReAct}: Synergizing Reasoning and Acting in Language Models},
  author = {Yao, Shunyu and Zhao, Jeffrey and Yu, Dian and Du, Nan and Shafran, Izhak and Narasimhan, Karthik and Cao, Yuan},
  booktitle = {ICLR},
  year = {2023},
  html = {https://arxiv.org/abs/2210.03629},
}

@article{weng2023agent,
  title   = "LLM-powered Autonomous Agents",
  author  = "Weng, Lilian",
  journal = "lilianweng.github.io",
  year    = "2023",
  month   = "Jun",
  url     = "https://lilianweng.github.io/posts/2023-06-23-agent/"
}

@misc{shen2025mindmachinerisemanus,
      title={From Mind to Machine: The Rise of Manus AI as a Fully Autonomous Digital Agent}, 
      author={Minjie Shen and Yanshu Li and Lulu Chen and Qikai Yang},
      year={2025},
      eprint={2505.02024},
      journal={arXiv},
      primaryClass={cs.AI},
      url={https://arxiv.org/abs/2505.02024}, 
}

@misc{xagent2023,
      title={XAgent: An Autonomous Agent for Complex Task Solving}, 
      author={XAgent Team},
      journal={arXiv},
      year={2023},
}

@Misc{smolagents,
  title =        {`smolagents`: a smol library to build great agentic systems.},
  author =       {Aymeric Roucher and Albert Villanova del Moral and Thomas Wolf and Leandro von Werra and Erik Kaunismäki},
  howpublished = {\url{https://github.com/huggingface/smolagents}},
  year =         {2025}
}

@article{shekhar2024optimizingcostsllmusage,
      title={Towards Optimizing the Costs of LLM Usage}, 
      author={Shivanshu Shekhar and Tanishq Dubey and Koyel Mukherjee and Apoorv Saxena and Atharv Tyagi and Nishanth Kotla},
      year={2024},
      eprint={2402.01742},
      journal={arXiv},
      primaryClass={cs.CL},
      url={https://arxiv.org/abs/2402.01742}, 
}

@misc{chatdbglm,
    author       = { {Chat2DB} },
    title        = { {Chat2DB-SQL-7B} },
    year         = 2024,
    url          = { https://huggingface.co/Chat2DB/Chat2DB-SQL-7B },
    journal    = { Hugging Face }
}

@article{russo2025abacuscostbasedoptimizersemantic,
      title={Abacus: A Cost-Based Optimizer for Semantic Operator Systems}, 
      author={Matthew Russo and Sivaprasad Sudhir and Gerardo Vitagliano and Chunwei Liu and Tim Kraska and Samuel Madden and Michael Cafarella},
      year={2025},
      eprint={2505.14661},
      journal={arXiv},
      primaryClass={cs.DB},
      url={https://arxiv.org/abs/2505.14661}, 
}

@article{dabstep,
      title={DABstep: Data Agent Benchmark for Multi-step Reasoning}, 
      author={Alex Egg and Martin Iglesias Goyanes and Friso Kingma and Andreu Mora and Leandro von Werra and Thomas Wolf},
      year={2025},
      eprint={2506.23719},
      journal={arXiv},
      primaryClass={cs.LG},
      url={https://arxiv.org/abs/2506.23719}, 
}

@article{deng2025reforcetexttosqlagentselfrefinement,
      title={ReFoRCE: A Text-to-SQL Agent with Self-Refinement, Consensus Enforcement, and Column Exploration}, 
      author={Minghang Deng and Ashwin Ramachandran and Canwen Xu and Lanxiang Hu and Zhewei Yao and Anupam Datta and Hao Zhang},
      year={2025},
      eprint={2502.00675},
      journal={arXiv},
      primaryClass={cs.CL},
      url={https://arxiv.org/abs/2502.00675}, 
}

@article{hu2024infiagentdabenchevaluatingagentsdata,
      title={InfiAgent-DABench: Evaluating Agents on Data Analysis Tasks}, 
      author={Xueyu Hu and Ziyu Zhao and Shuang Wei and Ziwei Chai and Qianli Ma and Guoyin Wang and Xuwu Wang and Jing Su and Jingjing Xu and Ming Zhu and Yao Cheng and Jianbo Yuan and Jiwei Li and Kun Kuang and Yang Yang and Hongxia Yang and Fei Wu},
      year={2024},
      eprint={2401.05507},
      journal={arXiv},
      primaryClass={cs.CL},
      url={https://arxiv.org/abs/2401.05507}, 
}

@article{lei2025spider20evaluatinglanguage,
      title={Spider 2.0: Evaluating Language Models on Real-World Enterprise Text-to-SQL Workflows}, 
      author={Fangyu Lei and Jixuan Chen and Yuxiao Ye and Ruisheng Cao and Dongchan Shin and Hongjin Su and Zhaoqing Suo and Hongcheng Gao and Wenjing Hu and Pengcheng Yin and Victor Zhong and Caiming Xiong and Ruoxi Sun and Qian Liu and Sida Wang and Tao Yu},
      year={2025},
      eprint={2411.07763},
      journal={arXiv},
      primaryClass={cs.CL},
      url={https://arxiv.org/abs/2411.07763}, 
}

@article{zhang2024codeagentenhancingcodegeneration,
      title={CodeAgent: Enhancing Code Generation with Tool-Integrated Agent Systems for Real-World Repo-level Coding Challenges}, 
      author={Kechi Zhang and Jia Li and Ge Li and Xianjie Shi and Zhi Jin},
      year={2024},
      eprint={2401.07339},
      journal={arXiv},
      primaryClass={cs.SE},
      url={https://arxiv.org/abs/2401.07339}, 
}

@misc{wang2024mineruopensourcesolutionprecise,
      title={MinerU: An Open-Source Solution for Precise Document Content Extraction}, 
      author={Bin Wang and Chao Xu and Xiaomeng Zhao and Linke Ouyang and Fan Wu and Zhiyuan Zhao and Rui Xu and Kaiwen Liu and Yuan Qu and Fukai Shang and Bo Zhang and Liqun Wei and Zhihao Sui and Wei Li and Botian Shi and Yu Qiao and Dahua Lin and Conghui He},
      year={2024},
      eprint={2409.18839},
      journal={arXiv},
      primaryClass={cs.CV},
      url={https://arxiv.org/abs/2409.18839}, 
}

@misc{chhikara2025mem0buildingproductionreadyai,
      title={Mem0: Building Production-Ready AI Agents with Scalable Long-Term Memory}, 
      author={Prateek Chhikara and Dev Khant and Saket Aryan and Taranjeet Singh and Deshraj Yadav},
      year={2025},
      eprint={2504.19413},
      journal={arXiv},
      primaryClass={cs.CL},
      url={https://arxiv.org/abs/2504.19413}, 
}

@misc{xu2025amemagenticmemoryllm,
      title={A-MEM: Agentic Memory for LLM Agents}, 
      author={Wujiang Xu and Zujie Liang and Kai Mei and Hang Gao and Juntao Tan and Yongfeng Zhang},
      year={2025},
      eprint={2502.12110},
      journal={arXiv},
      primaryClass={cs.CL},
      url={https://arxiv.org/abs/2502.12110}, 
}

@article{10.1145/3604931,
author = {Ehrmann, Maud and Hamdi, Ahmed and Pontes, Elvys Linhares and Romanello, Matteo and Doucet, Antoine},
title = {Named Entity Recognition and Classification in Historical Documents: A Survey},
year = {2023},
issue_date = {February 2024},
publisher = {Association for Computing Machinery},
address = {New York, NY, USA},
volume = {56},
number = {2},
issn = {0360-0300},
url = {https://doi.org/10.1145/3604931},
doi = {10.1145/3604931},
journal = {ACM Comput. Surv.},
month = sep,
articleno = {27},
numpages = {47}
}

@misc{zhang2024surveymemorymechanismlarge,
      title={A Survey on the Memory Mechanism of Large Language Model based Agents}, 
      author={Zeyu Zhang and Xiaohe Bo and Chen Ma and Rui Li and Xu Chen and Quanyu Dai and Jieming Zhu and Zhenhua Dong and Ji-Rong Wen},
      year={2024},
      eprint={2404.13501},
      journal={arXiv},
      primaryClass={cs.AI},
      url={https://arxiv.org/abs/2404.13501}, 
}

\end{document}